\pdfoutput=1
\documentclass{aa}

\usepackage{graphicx}
\usepackage{booktabs}
\usepackage{multirow}
\usepackage{color}  
\usepackage{txfonts}
\usepackage{subcaption}
\usepackage{float}
\usepackage{enumitem}
\usepackage[normalem]{ulem} 
\begin{document} 

\authorrunning{I.C. Jebaraj, et al.,}
\titlerunning{Interplanetary type III fine structures}

   \title{Structured type III radio bursts observed in interplanetary space}

   \author{Immanuel. C. Jebaraj
          \inst{1, 2},
          J. Magdalenic\inst{1,2},
          V. Krasnoselskikh\inst{3,4},
          V. Krupar\inst{5,6}
          and S. Poedts\inst{2,7}
          }
    \institute{1. Solar-Terrestrial Centre of Excellence -- SIDC, Royal Observatory of Belgium, Avenue Circulaire 3, 1180 Uccle, Belgium \\
              2. Center for mathematical Plasma Astrophysics, Department of Mathematics, KU Leuven, Celestijnenlaan 200B, B-3001 Leuven, Belgium \\
              3. LPC2E/CNRS, UMR 7328, 3A Avenue de la Recherche Scientifique, Orleans, France \\
              4. Space Sciences Laboratory, University of California, Berkeley, CA, USA \\
              5. Goddard Planetary Heliophysics Institute, University of Maryland, Baltimore County, Baltimore, MD 21250, USA \\
              6. Heliospheric Physics Laboratory, Heliophysics Division, NASA Goddard Space Flight Center, Greenbelt, MD 20771, USA \\
              7. Institute of Physics, University of Maria Curie-Sk{\l}odowska, ul.\ Radziszewskiego 10, PL-20-031 Lublin, Poland
             }
  
   \date{}

 
  \abstract
   {The last few decades has seen numerous studies dedicated to fine structures of type III radio bursts observed in the metric to decametric wavelengths. Majority of explanations of the structured radio emission involve the propagation of electron beam through the strongly inhomogeneous plasma in the low corona. Until now only few studies of single type III bursts with fine structures, observed in the hecto-kilometric wavelengths, were reported.}
   {Herein we report about existence of numerous structured type III radio bursts observed during the \textit{STEREO} era by all three \textit{WAVES} instruments on board \textit{STEREO A}, \textit{B}, and \textit{Wind}. The aim of the study is to report, classify structured type III bursts, and present the characteristics of their fine structures. The final goal is to try to understand the physical mechanism responsible for the generation of structured radio emission.}
   {In this study we used data from all available spacecraft, specifically the \textit{STEREO} and the \textit{Wind} spacecraft. We employ 1D density models to obtain the speed of the source of type III radio emission, the electron beam. We also perform spectral analysis of the fine structures in order to compare their characteristics with the metric-decametric fine structures. }
   {The presented similarities of the type III fine structures in the metric to decametric and interplanetary wavelengths indicate that the physical processes responsible for the generation of structured type III radio bursts could be the same, at the heights, all the way from the low corona to the interplanetary range. We show that the observed structuring and intermittent nature of the type III bursts can be explained by the variation in the level of density fluctuations, at different distances from the Sun.
   }
   {}


   \maketitle
%
\section{Introduction} \label{Sec:Introduction}

Type III radio bursts are among the most intense radio emissions of solar origin. They are also frequently observed radio bursts, by both space-based and ground-based radio observatories. They are the radio signatures of suprathermal electron beams propagating along open and quasi-open magnetic field lines. In a dynamic radio spectra (colour coded frequency-time diagrams) type III bursts are recorded as fast drifting bursts appearing at a wide range of frequencies almost simultaneously \citep[see e.g.][for review]{Suzuki85book}. Along with type II radio bursts which are the radio signatures of shock waves propagating in the solar corona, type III radio bursts are the most studied solar radio bursts because they can provide information about the associated eruptive event and the ambient plasma conditions. 

Solar radio bursts are observed at decreasing frequencies as the associated radio source propagates away from the Sun. In the metric to decametric wavelengths (300--20 MHz), radio emission can be studied using both, the dynamic spectra and interferometric observations \citep[see e.g.][]{Magdalenic10, Zucca14, Kontar17b, Zucca18, Sharykin18}. These two types of observations provide complementary information about the radio bursts, i.e. the shape and extent of radio bursts and the position of associated radio sources projected in the plane-of-the-sky. Due to the ionospheric cutoff \citep{Erickson97}, radio emission beyond $\approx 20$~MHz are only observed by some ground based instruments \citep[e.g.][]{Melnik18}.

Space based radio observations start at frequencies of about 10~MHz, and radio bursts observed below 3 MHz (hectometric to kilometer wavelengths) are considered to be interplanetary radio emission. Similar to the radio bursts in the metric range, interplanetary radio bursts are also mostly generated by propagating shock waves (type II bursts) and electron beams traveling along open field lines (type III bursts). Additionally, co-rotating stream interaction regions (SIR/CIR), or interaction between CMEs can result in the complex continuum-like emission \citep[e.g.][]{Gopalswamy01}.

Interplanetary radio emission of solar origin is generally considered to be generated by plasma emission mechanism. Bursts are observed at the fundamental of local plasma frequency, its second harmonic, and sometimes at both fundamental and second harmonic. Although the fundamental emission is dominant, the harmonic plasma emission is also sometimes observed in interplanetary space \citep[e.g.][]{Leblanc98}. 

Radio observations in the hectometric to kilometer wavelengths are mostly limited to the dynamic spectra which show the intensity of emission as a function of frequency and time, but lack the spatial information. Sometimes also the stereoscopic direction-finding observations are available. In particular, since the launch of \textit{STEREO} mission, direction-finding observations are continuously provided bringing us the unique opportunity to study the positions of radio sources and better understand origin of interplanetary radio emission \citep[see e.g.][]{MartinezOliveros12, Magdalenic14, Jebaraj20, Jebaraj21}.

Recent advancement in radio imaging techniques and space-based spectroscopic observations have improved the possibility for identification, categorization, and analysis of radio fine structures. Extensive reporting and studies are available for metric fine structure \citep[see for e.g.][]{Bhonsle79, Magdalenic06, Magdalenic20} of different types of radio bursts (type II, III, IV). On the contrary, there is a very small number of studies reporting the low frequency radio fine structures \citep[e.g.][]{Thejappa19, Thejappa20, Pulupa20, ChenL21}. Simultaneous observations from different viewing perspective provided by the \textit{STEREO} and \textit{Wind} spacecraft over the last decade, give us a large data base of long wavelength observations revealing number of different radio fine structures of both type II and type III radio bursts. 

In this study, we present for the first time, different radio fine structures observed by the \textit{Wind/WAVES} and \textit{STEREO/WAVES} instruments. We describe the characteristics and the morphology of interplanetary fine structures and discuss their possible origin.
Unfortunately it was not possible to perform radio triangulation study \citep[e.g.][]{Magdalenic14, Jebaraj20} to locate the position of the associated radio sources. Namely, majority of well observed structured type III bursts was recorded by spacecraft separated about 180 degrees which is not favourable for radio triangulation \citep{Krupar16}.

This paper is structured as follows: We describe the data and methods used in this analysis in Sec. \ref{Data}, and a brief introduction to interplanetary type III radio emission in Sec. \ref{IP_radio_typeIII}. We then describe the morphological characteristics of three different classes of interplanetary type III fine structures, namely, interplanetary type IIIb bursts (Sec. \ref{IP_typeIIIb_bursts}), type III bursts with triangular substructures (Sec. \ref{2_triangular}), and type III bursts with irregular fine structures (Sec. \ref{3_irregular}). In Sec. \ref{Sec:model}, we propose a model for the generation of fine structures and their characteristics. Finally, the summary of our findings and a discussion of the results can be found in Sec. \ref{Discussions}. 

\section{Data and Methods}\label{Data}
\subsection{Observations}\label{Observations}

In this study we used observations from \textit{WAVES} instruments onboard the twin Solar TErrestrial RElations Observatory Ahead and Behind \citep[\textit{STEREO} \textit{A} and \textit{B}][]{Kaiser05, Kaiser08, Bougeret08} spacecrafts and the \textit{Wind} spacecraft \citep[][]{Bougeret95}. All three \textit{WAVES} instruments provide intensity-time data at unique frequencies in the form of a radio dynamic spectra. The frequency range we utilise from the \textit{STEREO/WAVES} instrument is 125--16000~kHz (HFR receiver only), and from \textit{Wind/WAVES} is 4--13825~kHz (RAD~1 and RAD~2 receivers).

We started study by inspecting radio observations and when radio events were identified we looked for the possibly associated CME/flare events observed in the low corona. For that we used the following observations:

\begin{itemize}
    \item[\textbullet]
    The white-light (WL) coronagraph observations from the three vantage points: a) the Large Angle and Spectroscopic Coronagraph \citep[\textit{LASCO};][]{Brueckner95} on board the Solar and Heliospheric Observatory \citep[\textit{SOHO};][]{Domingo95}; and  b) the  coronagraphs on board \textit{STEREO A} \& \textit{STEREO B} \citep[][]{Kaiser08, Howard08}. 
    
    \item[\textbullet]
    The Extreme ultra violet (EUV) wavelengths observations provided by the Atmospheric Imaging Assembly \citep[\textit{AIA};][]{Lemen12} onboard Solar Dynamics Observatory \citep[\textit{SDO};][]{Pesnell12} (\textit{SDO/AIA} for short), and the Extreme Ultra Violet Imagers \citep[\textit{EUVI}; ][]{Howard08} instrument on-board \textit{STEREO} (\textit{STEREO/EUVI}).

\end{itemize}

 \begin{figure*}[h]
  \centering
  \begin{subfigure}[b]{0.55\textwidth}
    \includegraphics[width=\textwidth]{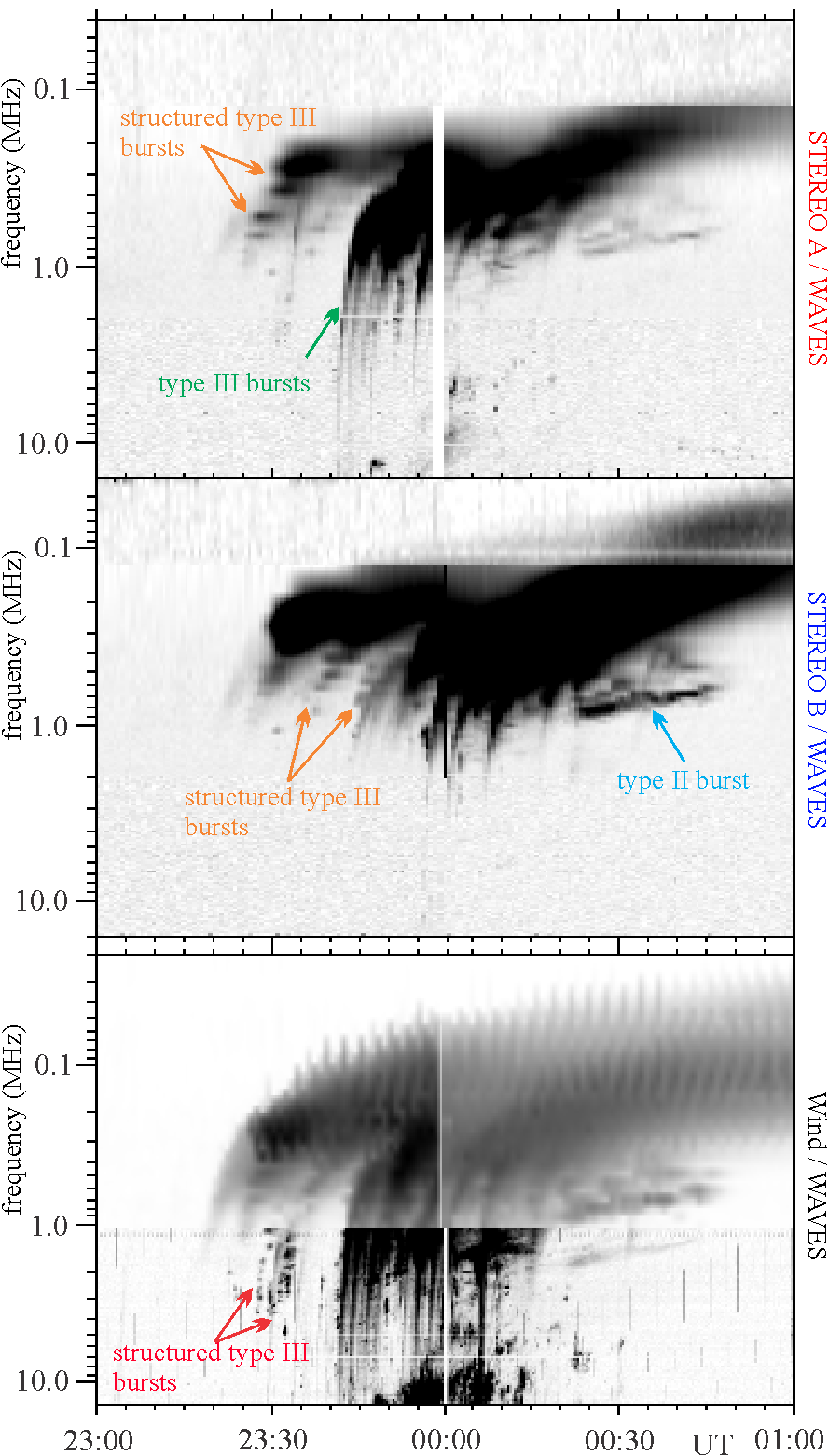}
  \end{subfigure}
  \caption{Complex radio event observed in the evening of September 27 and early morning of 28, 2012 shows type II burst, 'ordinary' interplanetary type III bursts, and structured type III bursts. Two different types of structured type III bursts are marked with different colors (orange and pink).}
  \label{Fig:example_spectra}
  \end{figure*}

\begin{table*}[]
\centering
\caption{Three classes of type III fine structures observed in interplanetary space. Figures showing examples of the studied classes of bursts are also marked in the table.}
\label{tab:my_table}
\resizebox{\textwidth}{!}{%
\begin{tabular}{@{}cccccc@{}}
\toprule
\hline
\textbf{Type of Fine structure} &
  \textbf{Date} &
  \textbf{Time} &
  \textbf{\begin{tabular}[c]{@{}c@{}}Bandwidth\\ (kHz)\end{tabular}} &
  \textbf{\begin{tabular}[c]{@{}c@{}}Duration\\ (seconds)\end{tabular}} &
  \textbf{Comment} \\ \midrule
\multirow{6}{*}{\begin{tabular}[c]{@{}c@{}}IP Type IIIb \\ bursts (Fig. \ref{Fig: figure_2}, Fig. \ref{Fig:figure_3})\end{tabular}} &
  19/09/2011 &
  \begin{tabular}[c]{@{}c@{}}06:45,\\  11:10\end{tabular} &
  60--90 &
  240--600 &
  \multirow{6}{*}{\begin{tabular}[c]{@{}c@{}}Most commonly observed type III fine\\ structures in IP space.\\ \\ Morphological characteristics are \\ somewhat similar to decametric type IIIb.\\ \\ Exhibit clear structuring and are well separated.\end{tabular}} \\ \cmidrule(lr){2-5}
 & 21/11/2011 & \begin{tabular}[c]{@{}c@{}}13:30, \\ 15:00, \\ 16:00\end{tabular} & 70--100  & 300--600 &  \\ \cmidrule(lr){2-5}
 & 22/11/2011 & \begin{tabular}[c]{@{}c@{}}18:10, \\ 19:00, \\ 20:00\end{tabular} & 70--100  & 240--600 &  \\ \cmidrule(lr){2-5}
 & 23/11/2011 & \begin{tabular}[c]{@{}c@{}}10:00, \\ 11:00\end{tabular}           & 120--200 & 180--660 &  \\ \cmidrule(lr){2-5}
 & 11/12/2011 & 20:00                                                             & 60-90   & 150--450 &  \\ \cmidrule(lr){2-5}
 & 27/09/2012 & \begin{tabular}[c]{@{}c@{}}23:25,  \\ 23:25\end{tabular}          & 70--90   & 240--600 &  \\ \midrule
\begin{tabular}[c]{@{}c@{}}Triangular striae\\  burst (Fig. \ref{Fig:figure_6})\end{tabular} &
  12/11/2010 &
  07:55 &
  90--120 &
  30--300 &
  \begin{tabular}[c]{@{}c@{}}Rarest type III fine structures.\\ Can also be observed in the higher hectometric\\ frequencies.\end{tabular} \\ \midrule
\multirow{4}{*}{\begin{tabular}[c]{@{}c@{}}Irregular type IIIb\\ burst (Fig. \ref{Fig:figure_9})\end{tabular}} &
  12/11/2010 &
  13:45 &
  90 -- 120 &
  60 -- 300 &
  \multirow{4}{*}{\begin{tabular}[c]{@{}c@{}}Observed mostly in higher \\ hectometric wavelengths.\\ \\ Have no clear structuring and can occur \\ in irregular intervals.\end{tabular}} \\ \cmidrule(lr){2-5}
 & 31/07/2011 & 19:00                                                             & 50--90   & 120--160 &  \\ \cmidrule(lr){2-5}
 & 11/08/2011 & 10:35                                                             & 30--80   & 30--120  &  \\ \cmidrule(lr){2-5}
 & 22/11/2011 & 01:20                                                             & 60--100  & 120--300  &  \\ \cmidrule(lr){2-5}
 & 30/11/2011 & 07:50                                                             & 80--110 & 60--150  &  \\ \bottomrule
\end{tabular}%
}
\end{table*}

\section{Interplanetary type III radio bursts}\label{IP_radio_typeIII}

Out of five main types of plasma radio emissions of solar origin, type III radio bursts are the most frequently observed one. Type III radio bursts are generated by beams of suprathermal electrons (a few keVs to tens of keVs) propagating anti-sunward along open and quasi-open magnetic field lines. The broadband emission produced by fast electron beams is observed in dynamic spectra as rapidly drifting bursts \citep[see Fig.~\ref{Fig:example_spectra} and for a review see e.g.][and references therein]{Bhonsle79}. Based on their morphology different types of metric to decametric type III bursts were distinguished \cite[see e.g.][]{Kundu65, Bhonsle79}. Structured type III bursts were frequently reported, but only at metric and decametric \citep[e.g.][]{Chernov07, Melnik18}.

For the past 14 years of the \textit{STEREO} era numerous hectometric to kilometric type III radio bursts were observed, and some of them were also structured type III bursts. Herein, we first time report and classify the most frequently observed structured interplanetary type III bursts. 

We distinguish three main categories of type III bursts with fine structures: (i) interplanetary type IIIb bursts with and without an envelope, (ii) type III bursts with triangular substructures and (iii) type III bursts with irregular substructures. Table \ref{tab:my_table} lists different radio events, addressed in our study, with structured interplanetary type III bursts. We note that the presented events are only selected examples, and larger number of structured type III bursts was found.

 \begin{figure*}[h]
  \centering
  \begin{subfigure}[b]{0.75\textwidth}
    \includegraphics[width=\textwidth]{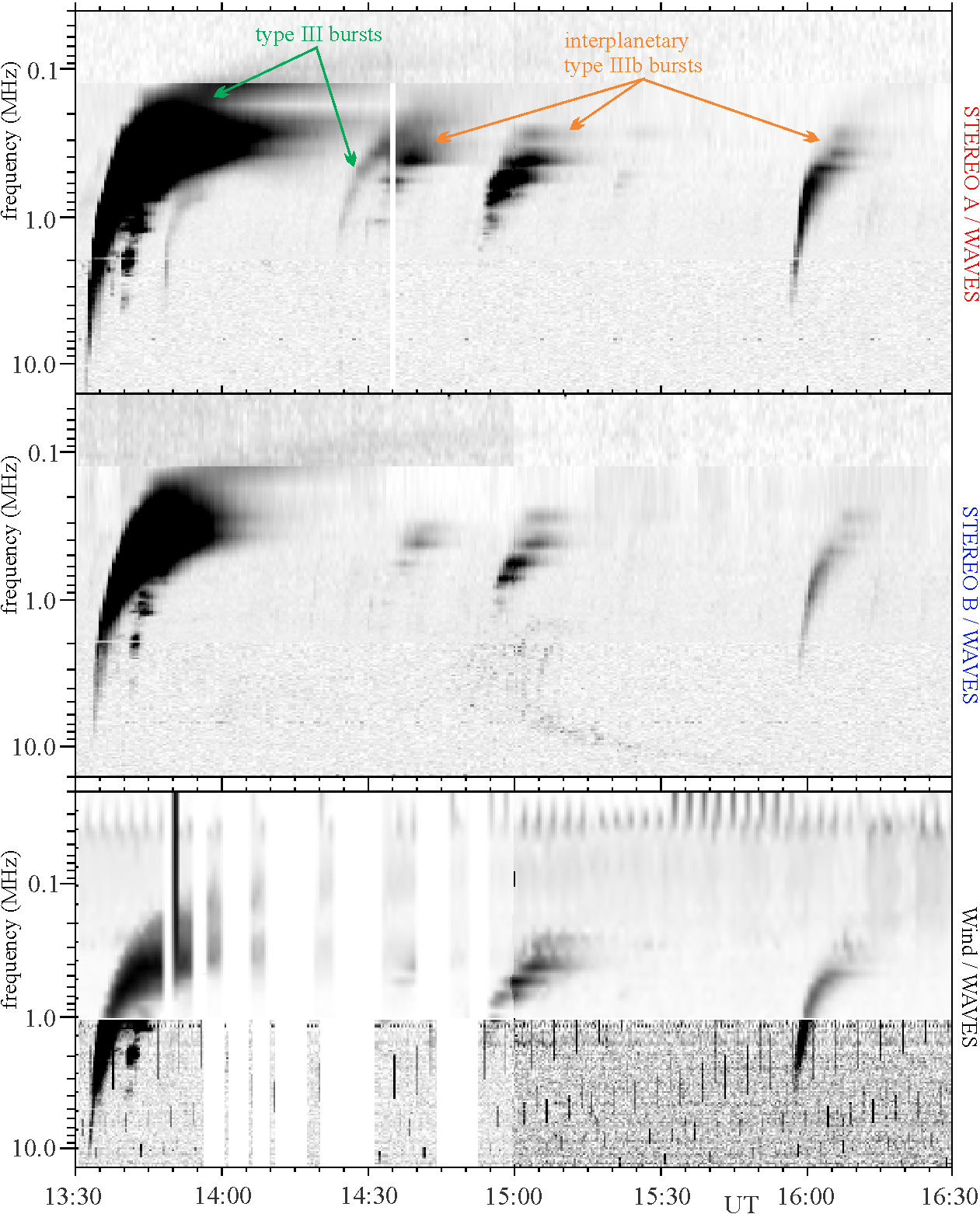}
  \end{subfigure}
  \caption{Example of few structured interplanetary Type III bursts observed on 21/11/2011. The three type IIIb bursts show striae substructures, mostly visible in frequencies below 700 kHz. The III burst with irregular substructures is visible at about 13:40 UT. This structured type III is possibly harmonic component of the strong type III burst observed shortly after 13:30 UT.}
  \label{Fig: figure_2}
  \end{figure*}

 \begin{figure*}[h]
  \centering
  \begin{subfigure}[b]{0.65\textwidth}
    \includegraphics[width=\textwidth]{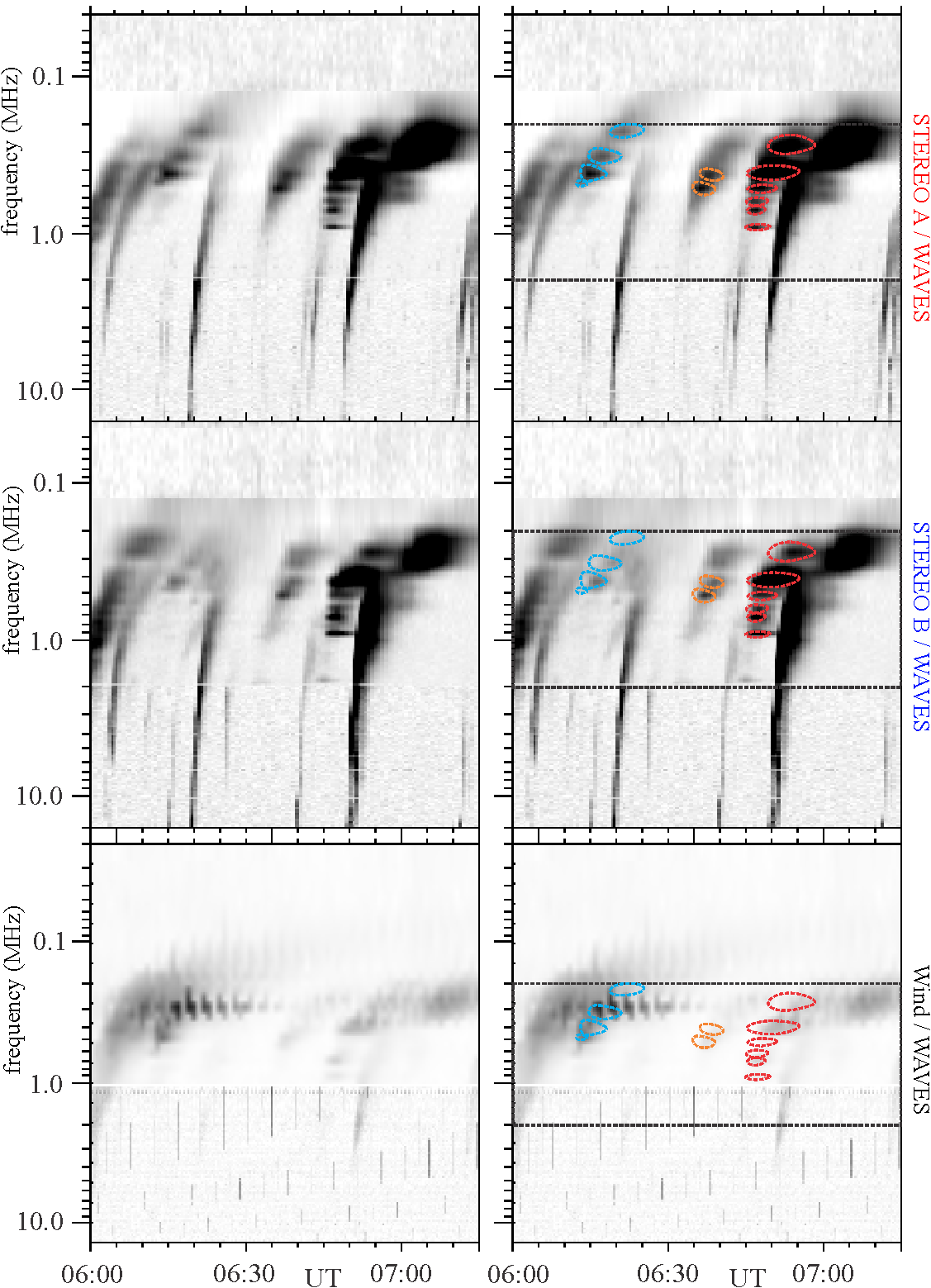}
  \end{subfigure}
  \caption{Radio Dynamic spectra observed on 19/09/2011 between 6:00 UT and 7:30 UT. (a) The IP type IIIb bursts consisting of striae substructures were observed along with multiple, regular type III radio bursts by \textit{STEREO A \& B}. (b) Schematic representation of the striae fine structures as observed by \textit{STEREO A} and overlaid on \textit{STEREO B} and \textit{Wind} observations. The frequency range (2000 -- 200 kHz) between which the IP type IIIb bursts were observed is marked by the black frame .}
  \label{Fig:figure_3}
  \end{figure*}

\subsection{Interplanetary type IIIb bursts} \label{IP_typeIIIb_bursts}

\subsubsection{General characteristics} \label{1_general_char}

The most frequently observed type III radio bursts with fine structures are the type IIIb bursts. This subcategory of bursts are morphologically very similar to the metric to decametric type IIIb bursts  \citep[first reported by][]{delaNoe72}. The interplanetary type IIIb radio bursts are sometimes associated with an eruptive CME/flare events and sometimes only with the complex active regions, similarly to ordinary type III bursts.
Although some of the type IIIb bursts are observed at frequencies of up to 1000 kHz they are primarily recorded at frequencies bellow 700 kHz and then they do not have a high frequency type III counterpart. Out of twelve radio events presented in the Table. \ref{tab:my_table}, six events consisted of the type IIIb bursts.

Despite the observed fine structures, the overall spectral morphology, i.e. the envelope of the interplanetary type IIIb bursts is similar to that of the classical type III bursts. The same characteristic was found also for the decameter type IIIb bursts \citep[][]{Ellis67, delaNoe72, ChenX18, Zhang20}.

The main characteristic of the interplanetary type IIIb bursts is their well defined substructures, the so called \textit{striae}. The striae substructures are generally organized within the type III burst envelope as shown in Figs. \ref{Fig:example_spectra}, \ref{Fig: figure_2}. They sometimes show quasi-periodicity (Figs. \ref{Fig:example_spectra}, \ref{Fig: figure_2}), and sometimes an irregular pattern (Fig. \ref{Fig:figure_3}). The frequency bandwidth of the striae was found to be in the range of 70--120 kHz. For comparison, the bandwidth of striae structures observed in the metric to decametric range varies between 30 and 300 kHz.

In a majority of events the duration of striae is frequency dependent, i.e. duration of these narrowband structures are longer at lower observing frequencies. However, the irregular striae structures, with longer duration at higher frequency, are also occasionally observed (Fig. \ref{Fig:figure_3}). Their duration is in average between 120 seconds (at 4000 kHz) and 600 seconds (at 200 kHz), i.e. five times longer duration at five times lower frequency. This regularity in the bandwidth and duration suggests similar plasma characteristics necessary for the generation of interplanetary type IIIb bursts. 

The envelope of the type IIIb bursts is generally of significantly lower intensity than the substructures of the bursts. Therefore, in majority of studied radio events dynamic spectrum alone cannot clearly show if the type IIIb envelope exists or not. This is way we inspected also the individual frequency time profiles of the studied bursts (see section 3.1.1 and Fig. \ref{Fig:figure_5}).

 \begin{figure*}[h]
  \centering
  \begin{subfigure}[b]{0.6\textwidth}
    \includegraphics[width=\textwidth]{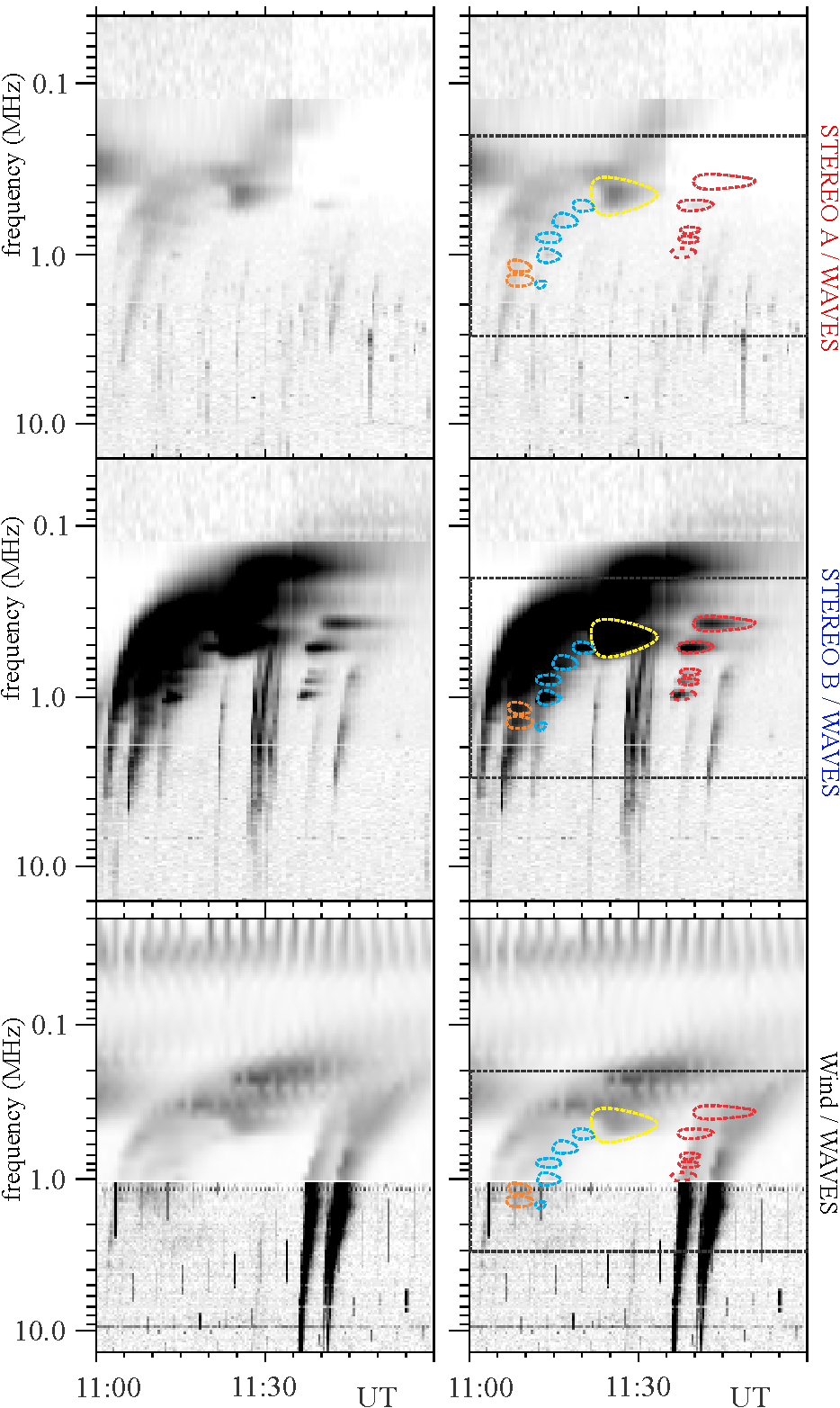}
  \end{subfigure}
  \caption{Radio Dynamic spectra observed on 19/09/2011 shows group of type III bursts. (a) Type III bursts with striae substructures were observed along with regular type III radio bursts. Emission was strongest in observations by \textit{STEREO B}. (b) schematic representation of the outlines of the striae structures as observed by \textit{STEREO B} and overlaid on \textit{STEREO A} and \textit{Wind} observations. The black frame in three panels represents the same frequency range of 3000-200 KHz.}
  \label{Fig:figure_4}
  \end{figure*}



\subsubsection{Examples of type IIIb bursts} \label{1_examples}

Figures \ref{Fig:example_spectra}, \ref{Fig: figure_2}, \ref{Fig:figure_3}, and \ref{Fig:figure_4} show different examples of the type IIIb bursts. The radio events presented in Figs. \ref{Fig: figure_2}, \ref{Fig:figure_3}, and \ref{Fig:figure_4} were associated with the repeated eruptive activity from the complex active regions (as observed in the EUV by either \textit{SOHO} or \textit{STEREO} spacecraft). Only the radio event observed on September 27, 2012 (Fig. \ref{Fig:example_spectra}) was clearly associated with the CME/flare event studied in details by \cite{Jebaraj20}. This radio event consists of numerous type III radio bursts, structured type III bursts and two type II bursts. The  structured interplanetary type III bursts were very clearly observed by both twin \textit{STEREO} spacecraft. The most prominent are IIIb radio bursts marked by the orange line in the Fig. \ref{Fig:example_spectra}. The structured radio emission was observed simultaneously by two different instruments and it is therefore certain that it is not an instrumental effect.

Fig. \ref{Fig: figure_2} shows several subsequent type III bursts, three of them type IIIb bursts, one type III burst with irregular substructures and few ordinary type III bursts. Two type IIIb bursts (at about 13:40 UT and 14:30 UT) could be the structured harmonic emission with the fundamental counterpart observed as regular type III. Such a pairs of type III bursts were already reported in the decameter range \citep[see e.g.][]{Melnik18,Melnik19, Zhang20}. The type IIIb bursts at about 15:00 UT and 16:00 UT do not show such a fundamental-harmonic pair as only structured type IIIb emission was observed. However, the bursts sometimes appear without specific sequence, i.e. we observe that the ordinary type III can be followed by the structured one or opposite. In such events it is difficult to clearly distinguish if they make type III-type IIIb pairs, similar to the decametric bursts \citep{Melnik18,Melnik19}. 

\subsubsection{Radio events on September 19, 2011} \label{1_case_study}

The radio events on September 19, 2011 exhibit particularly well observed striae structures of the type IIIb burst (Figs.\ref{Fig:figure_3} and \ref{Fig:figure_4}), and we discuss them herein in details.

Fig. \ref{Fig:figure_3} presents one more radio event on September 19, 2011, observed in the time interval 06:00 -- 07:15 UT. Similarly to the right hand panel of Fig. \ref{Fig:figure_4}, we overlaid the one-to-one schematic presentation of fine structures on the dynamic spectrum. The schematic presentation was done on the \textit{STEREO A} observations in which the bursts were best defined. Different colours, indicate fine structures belonging to different bursts. The starting frequency and the intensity of the striae is very similar, practically identical, in both \textit{STEREO A} and \textit{STEREO B} observations. Further, the morphology of striae structures seem to be almost identical as seen by the two \textit{STEREO} spacecraft. The one-to-one schematic presentation of striae (Fig. \ref{Fig:figure_3}, right panel) outlines these similarities in the spectral morphology. Taking into account both, morphology and timing of the structured type IIIb bursts, and assuming that the radio emission is brightest in the direction of its propagation \citep[e.g.][]{Magdalenic14, Jebaraj20}, we can approximate that the source location of the striae is more or less equidistant for both spacecraft.  




The \textit{Wind} observations show very faint fine structures with about two orders of magnitude lower intensity than the bursts observed by the twin \textit{STEREO} spacecraft. Due to low intensity of the bursts, it is not possible to estimate the exact start time of the striae structures. The detection of type IIIb burst in \textit{Wind} observations would be unreliable if not confirmed by \textit{STEREO} observations. What we can conclude from \textit{Wind} observations is that the structuring of the radio emission is not intensity dependent, and striae can be generated also within the very low intensity radio bursts. 


The frequency drift rate of the type IIIb burst, estimated along the brightest parts of the striae structures, is larger than that of the regular type III observed straight after (at about 06:50 UT). The frequency-drift rate of the type IIIb is 14 kHz s$^{-1}$ in the range 1000 -- 500 kHz \citep[which corresponds to $\approx 0.24$c using a 2-fold Leblanc density model][]{Leblanc98}, while that of the type III burst in the same frequency range is about 10 kHz s$^{-1}$ ($\approx 0.18$c). On the other hand the faint envelope of type IIIb shows very similar drift rate, of about 9 kHz s$^{-1}$ ($\approx 0.16$c), to the drift rate of the regular type III burst. A similar speed difference between the exciters of type IIIb and the regular type III was found in previous studies \citep[see e.g.][]{Baselyan74a, Baselyan74b, Sawant78, Melnik19} which suggested that the electron beams generating type IIIb were generally faster than the ones generating the regular type III bursts. It is important to note that such estimations of beam velocity from the spectral drift rate of type III radio bursts should only be used as first order approximation since they assume a 1D radial density model \citep[][]{Jebaraj20}. Furthermore, the electrons generating type III radio emissions are discrete and should be considered as a distribution of energies rather than being mono-energetic \citep[][]{Mann22a}.

\begin{figure*}[h]
  \centering
  \begin{subfigure}[b]{0.95\textwidth}
    \includegraphics[width=\textwidth]{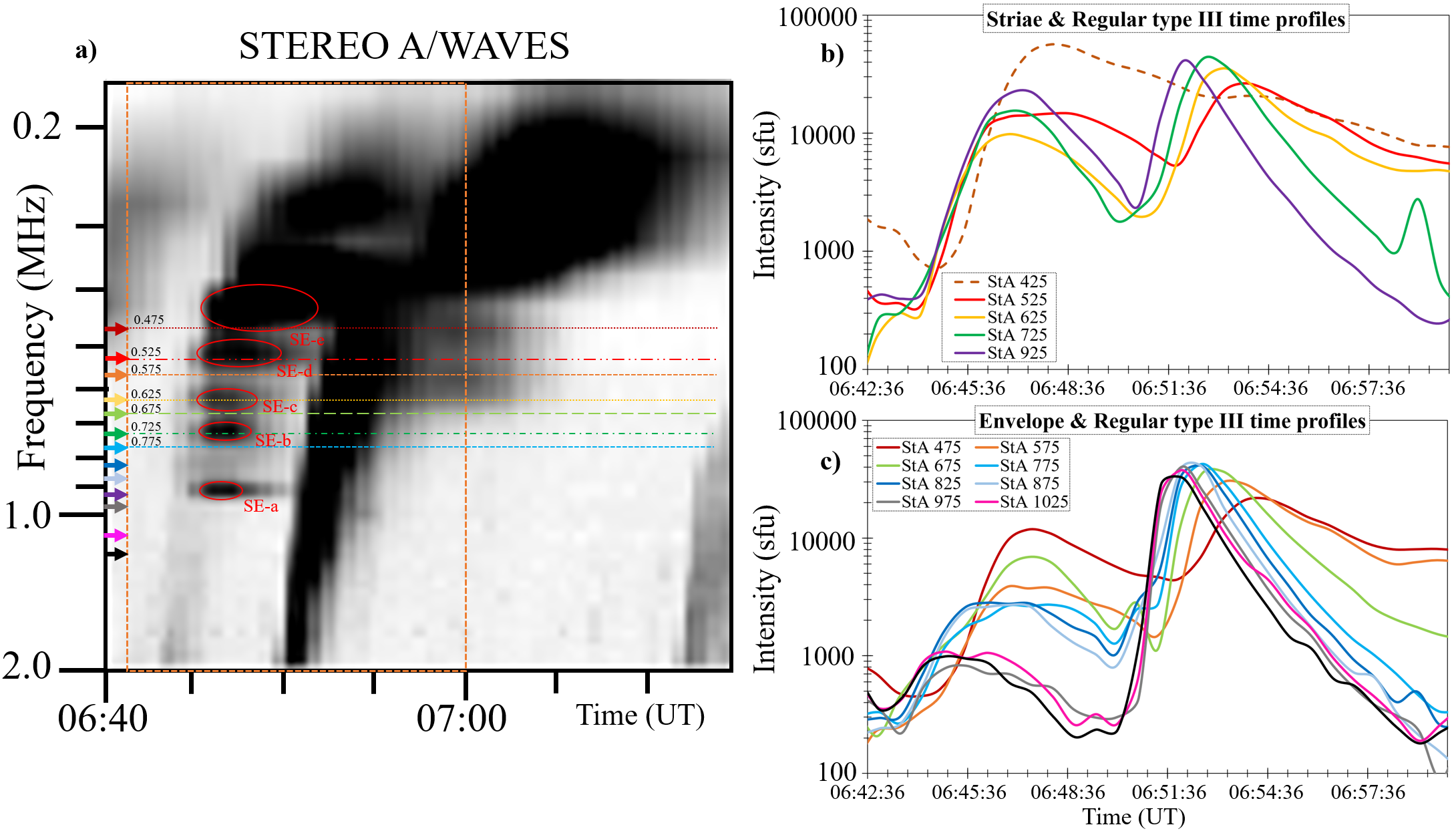}
  \end{subfigure}
  \caption{(a) Type IIIb bursts with striae like fine structures observed on 19/09/2011 by \textit{STEREO B} spacecraft. The structured type III is followed by very intense regular type III burst. (b) The time profiles of the type IIIb envelop and regular type III at selected frequencies. (c) The time profiles along the striae structures and regular type III burst. The selected frequencies (at panels b and c) are marked with the arrows of corresponding colours in the panel a. The time interval plotted in the panels b and c is within the orange box on panel a.}
  \label{Fig:figure_5}
  \end{figure*}

The properties of striae substructures within the type IIIb burst observed at 06:44 UT (Fig. \ref{Fig:figure_5}), are different in comparison to the striae bursts observed in the metric and decametric range. The duration of striae in this interplanetary type IIIb is longer than the 1/4 of the type III bursts duration, \citep[as reported by][in the decameter range]{Melnik19}. The bandwidth of the hectometric/kilometric striae presented here (about 70 -- 120 kHz ) is somewhat larger than in the decameter range (about 60 -- 80 kHz). We have found that the duration of striae can vary between 300 -- 600 s, which is typical duration of type III radio bursts in this frequency range.



Fig. \ref{Fig:figure_4} shows the radio bursts observed in the time interval 11:00 -- 12:00 UT were most intense in the \textit{STEREO B} observations ). The stochastic appearance of type IIIb bursts, recorded in radio event presented in Fig.\ref{Fig:figure_4} was also observed in the first radio event on September 19, 2011 (Fig. \ref{Fig:figure_3}). The right hand panel of the Fig. \ref{Fig:figure_4} shows the one-to-one schematic representations of the type III fine structures on the top of the dynamic spectrum. The one-to-one schematic presentation of the fine structures was done on the \textit{STEREO B} observations in which the bursts were best observed. The identical schematic drawing was overlaid on \textit{STEREO A} and \textit{Wind} observations, in order to compare the morphological characteristics of structured bursts, as observed by different spacecraft. We note that only a few, the most intense structures, were observed by all three spacecraft. Different colours indicate fine structures belonging most probably to different radio bursts. We note that the perfect temporal coincidence of the type III fine structures, as observed by two different spacecraft, should be expected only in the case when the electron beam that generates type III burst propagates exactly in the middle between the two spacecraft. Otherwise, the shift in the time profile of the fine structures will be observed, i.e. the signal will be earlier recorded by the spacecraft towards which the electron beam is propagating. Additionally, the intensity of the radio emission will be increasing or decreasing, depending if the electron beam propagates towards, or away from the spacecraft which records dynamic spectrum.

The striae substructures of two type IIIb bursts observed during two radio events on 19/09/2011 (Figs. \ref{Fig:figure_3} and \ref{Fig:figure_4})  have a lot of similarities. However, the organization of the striae substructures is more random in the second type IIIb, and the envelope of the burst is very faint. The striae structures are clearly observed only by \textit{STEREO B}, but the intensity of the radio emission seem to be larger in the \textit{Wind} observations. These radio event seem to be associated with an eruption from the NOAA AR11302. We found number of type IIIb bursts with similar striae-like substructures, few days earlier and later that September 19, 2011. 





 \begin{figure}[ht]
  \centering
  \begin{subfigure}[b]{0.49\textwidth}
    \includegraphics[width=\textwidth]{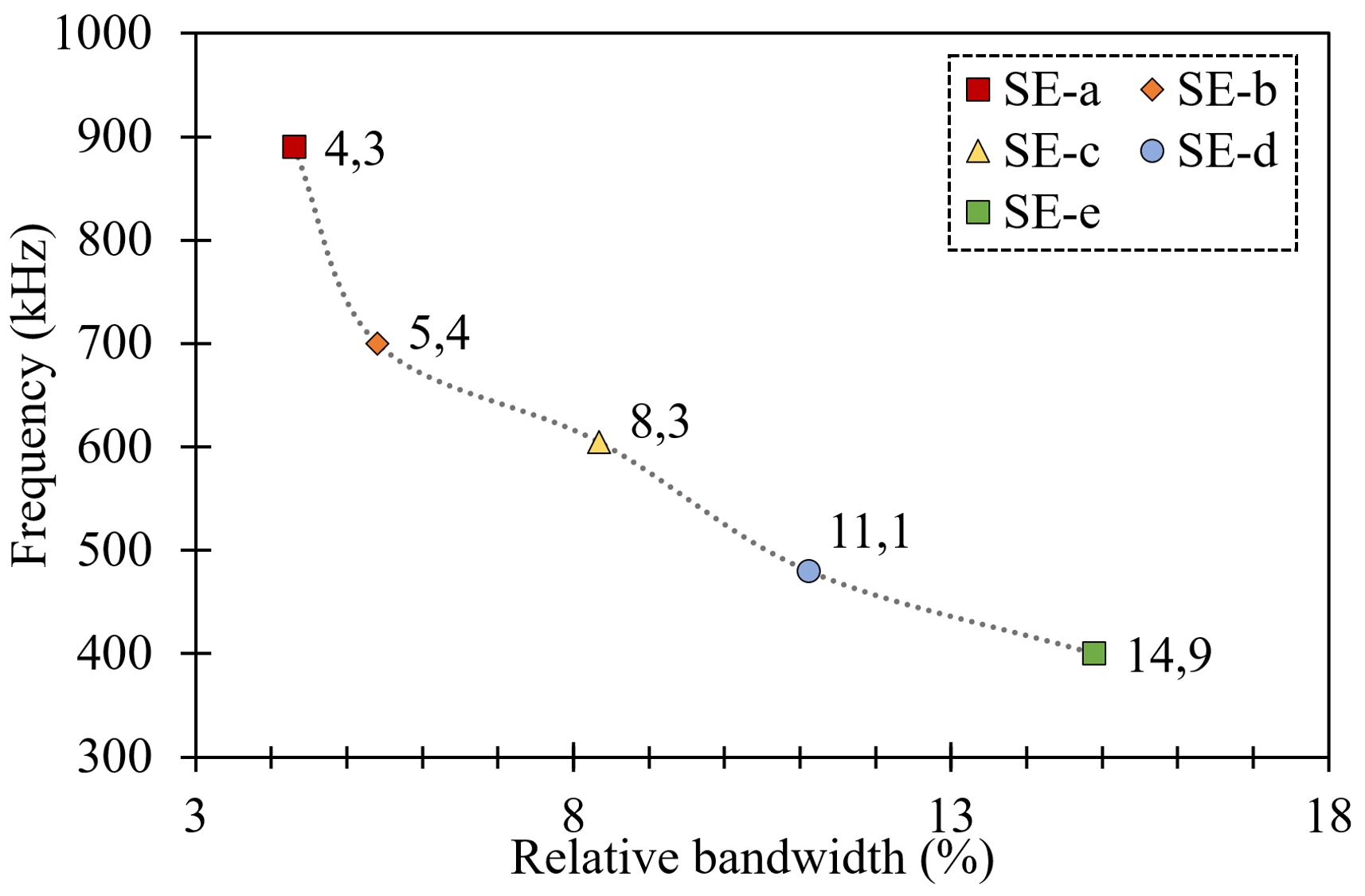}
  \end{subfigure}
  \caption{Relative bandwidth of the different striae elements represented by the abbreviations SE-a through SE-e. The values are presented as percentage of the central observing frequency.}
  \label{Fig:figure_rel_band}
  \end{figure}
  

\subsubsection{Time profiles for September 19, 2011 event} \label{1_time_profiles}

In this section we discuss the two subsequent type III bursts, the type IIIb burst with striae substructures and the regular type III burst which followed it. Figure \ref{Fig:figure_3} shows two bursts starting at about 06:44~UT and 06:50~UT, respectively. The  bright striae structures of type IIIb are observed in the frequency range of $1000-200$ kHz. The observed characteristics of type III bursts enabled qualitative interpretation of the dynamic properties of the substructures based on properties of beam plasma interaction in randomly in-homogeneous plasma \citep[][]{Krafft13, Krafft14, Krafft15, Voshchep15a, Voshchep15b, Krasnoselskikh19}.

 \begin{figure*}[h]
  \centering
  \begin{subfigure}[b]{0.99\textwidth}
    \includegraphics[width=\textwidth]{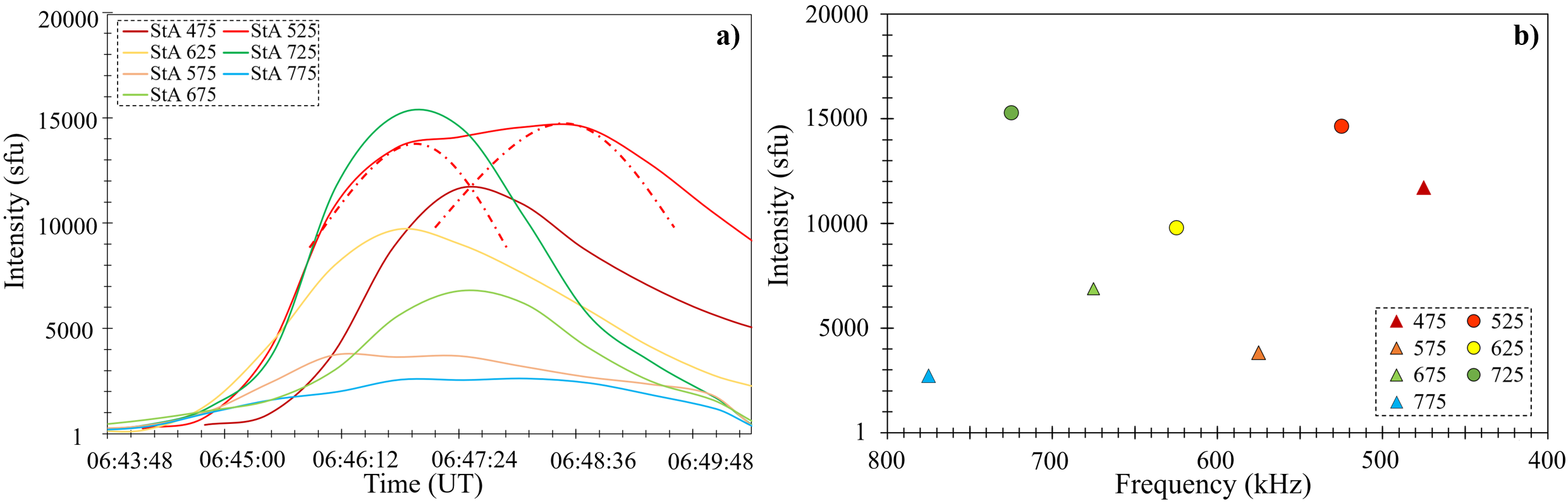}
  \end{subfigure}
  \caption{Spectral characteristics of the striae and envelope in the frequency range 800 to 450 kHz. Panel (a) shows the time profiles of the SE and their envelope in the given frequency range. The red dotted Gaussian-like curves represent the two peak morphology of the 525 kHz frequency band. Panel (b) shows the amplitude of the striae and envelope presented in panel (a). The triangle and circle markers represent the peak amplitude of the envelope and striae elements, respectively. }
  \label{Fig:striae_envelop_tf_peak}
  \end{figure*}



Figure \ref{Fig:figure_5}a presents a detail of the radio dynamic spectra recorded on September 19, 2011 (Fig. \ref{Fig:figure_3}). The colored arrows indicate the frequencies of the time profiles of the two observed bursts (Fig.~\ref{Fig:figure_5} panels b and c).
The dynamic spectrum clearly shows that the type IIIb consists of the faint envelop emission and intense striae structures. The time profiles at few selected frequencies, crossing mostly through the central part of striae (marked as SE-a,b,c,d, and e), are presented in Fig. \ref{Fig:figure_5}a. The time profiles of the envelope (Fig. \ref{Fig:figure_5}c) have noticeable lower intensity that the time profiles of striae, and the difference can be as large as order of the magnitude.

Type III radio bursts often have a Gaussian-like rise profile and a power-law decay profile, with the rise time being  significantly shorter than the decay time \citep[e.g.][]{Reid14Review}, however an exponential decay profile has also been recently reported by \cite{Krupar18} and \cite{Musset21}. This characteristic is also clearly visible in the profiles presented in Fig. \ref{Fig:figure_5}. The peaks of the type III profiles show a systematic shift in time (higher to lower frequency), indicating the propagation of the electron beam and decrease of the ambient solar wind density. This characteristic may not be so clearly seen in the case of the striae structures. However, we do observe the increase of relative bandwidth and striae's duration with the decrease of observing frequency. Figure \ref{Fig:figure_rel_band} shows the change of the relative bandwidth of striae, presented in percents of the central frequency. This behaviour results from the angular broadening of the electron beam (caused by diverging magnetic field lines), as the beam propagates away from the Sun and the region occupied by the beam at each instance increases. A consequence of the increasing area of the beam is that the macroscopic in-homogeneity of the solar wind density increases within the area covered by the beam.



Figure \ref{Fig:striae_envelop_tf_peak}a shows the comparison of the characteristics of the striae and the envelope in the frequency range 450--800 kHz. Time profiles at 525 kHz, 625 kHz, and 725 kHz cross the central part of the striae and other profiles correspond to the envelope. The profiles show an exponential growth up to the maximum of the time profile and then a significantly slower decay. The growth rate of the time profile at 725 kHz, which is at the centre of the striae element, is significantly larger than the growth rate for frequencies 775 kHz and 675 kHz which cross the envelope on the lower and higher frequency side of the striae. A similar behaviour was found for the time profile of the striae element at 625 kHz, and the closest envelope profiles at 675 kHz and 575 kHz. The time profile at the 525 kHz, which is close to the central frequency of striae SE-d is somewhat different. Instead of a single maxima we observe a kind of plateau which might result from combination of two Gaussian-like profiles similar to those at 725 and 625 kHz that follow one after another and overlap (marked by the dotted red curves in Fig. \ref{Fig:striae_envelop_tf_peak}a). The neighboring 575 kHz profile which passes through the envelope, has a significantly smaller growth rate and a factor of four smaller amplitude. On the other hand, the time profile at 475 kHz passing through lower frequency envelope shows a higher amplitude than the time profiles of other envelope regions. This is because the envelope region through which the 475 kHz time profile is crossing is close to and partially includes some higher frequency parts of striae SE-e. A comparison of the maximum amplitude of these spectral lines show the very same ordering (Fig. \ref{Fig:striae_envelop_tf_peak}b).

 \begin{figure*}[h]
  \centering
  \begin{subfigure}[b]{0.6\textwidth}
    \includegraphics[width=\textwidth]{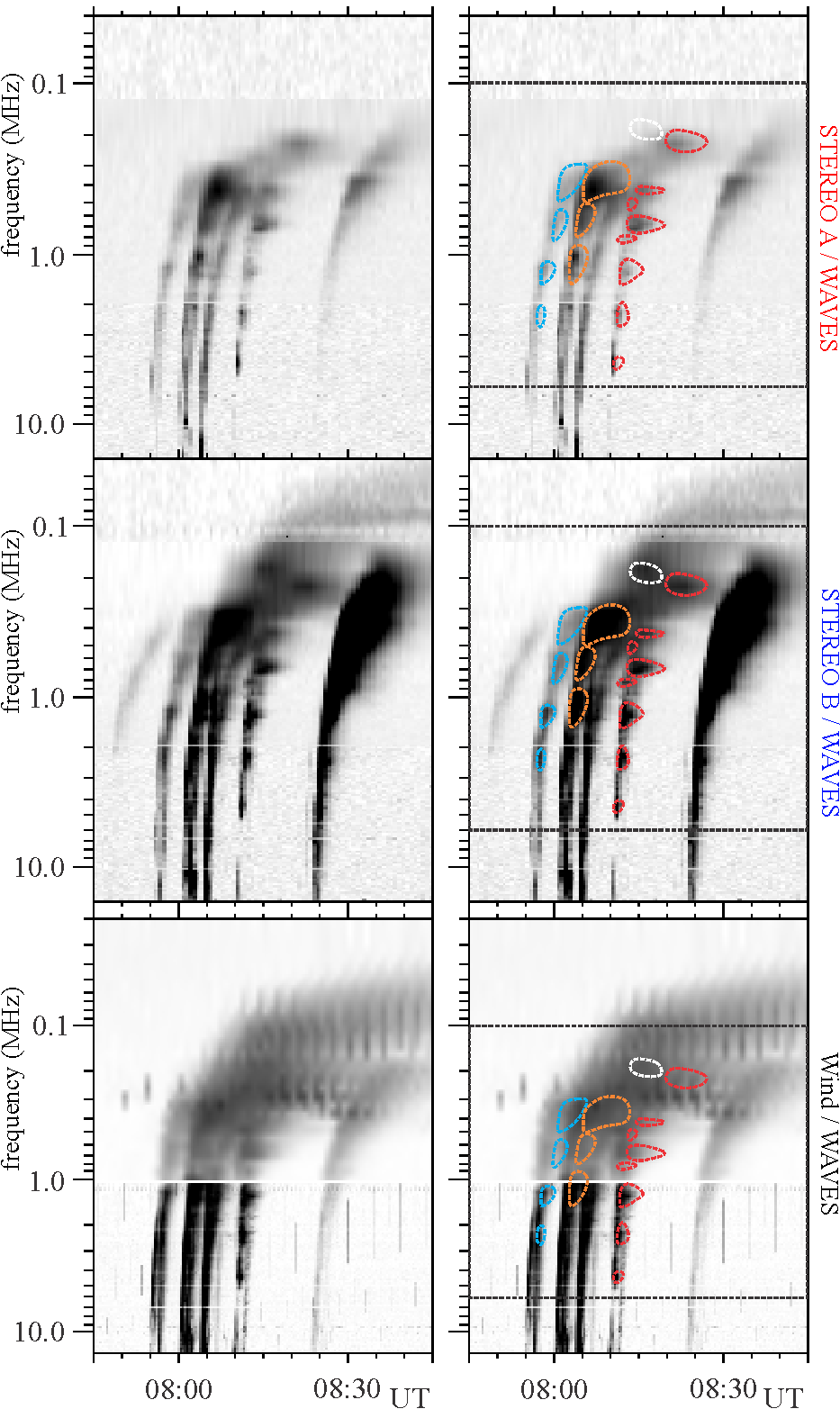}
  \end{subfigure}
  \caption {Radio Dynamic spectra observed on 12/11/2010. (a) Triangular striae substructures observed along with regular type III radio bursts by all three spacecraft (\textit{STEREO A} \& \textit{B}, and \textit{Wind}) at about 08:10 UT. (b) schematic representation of the outlines of the triangular striae structures as observed by \textit{STEREO B} and overlaid on \textit{STEREO A} and \textit{Wind} observations. Spectra also shows other regular type III radio bursts during the same time period.}
  \label{Fig:figure_6}
  \end{figure*}

\subsection{Type III bursts with triangular fine structures} \label{2_triangular}

\subsubsection{General characteristics} \label{2_general_char}

The type III bursts with triangular fine structures are second class of the structured interplanetary type III bursts. This type of bursts is observed rarely, in comparison with the interplanetary type IIIb bursts. Fig. \ref{Fig:figure_6} shows one group of these structured bursts. The spectral morphology of type III bursts with triangular fine structures is visibly different from the morphology of the type IIIb bursts presented in the previous section. The triangular substructures can be periodical, within the type III envelope, and they often partially overlap. Therefore, it is frequently difficult to distinguish between the fine structures and the underlying envelop. As it can be seen from Fig. \ref{Fig:figure_6}, the starting frequency of triangular substructures is generally higher than in the case of type IIIb bursts, going even up to 5~MHz.  

We have found that bandwidth of triangular substructures is generally between 90 and 120 kHz, which is larger than in a case of striae substructures. The full duration, estimated at the tip of the triangle, i.e. longest part, at 5 MHz is about 60 seconds, and 300 seconds at 200~kHz. The occurrence of the triangular substructures within the type III envelope decreases with frequency. Further, it seems that the triangular fine structures are composed of very narrow-band striae substructures (Fig. \ref{Fig:figure_8}), which are at the limit of the both, frequency and time resolution of the \textit{WAVES} observations.

\subsubsection{Radio event on November 12, 2010} \label{2_case_study}

Fig. \ref{Fig:figure_6} shows 5 different type III bursts observed by all three spacecraft on 12/11/2010. All type III bursts, except the one observed at about 08:25~UT, are structured. We observe significant difference between morphology and bandwidths of the triangular fine structures and the striae bursts, as discussed in the previous section.

The one-to-one schematic of the type III fine structure can be seen in Fig. \ref{Fig:figure_6}b. The bursts are the most intense in \textit{STEREO B} observations, therefore we use it as a reference for the comparison of morphological characteristics with \textit{STEREO A} and \textit{Wind}. The comparison of observations by three different spacecraft show very good match between all the fine structures, confirming that they are not due to the receiver artifact, or some technical problem. 

\begin{figure*}[h]
  \centering
  \begin{subfigure}[b]{0.99\textwidth}
    \includegraphics[width=\textwidth]{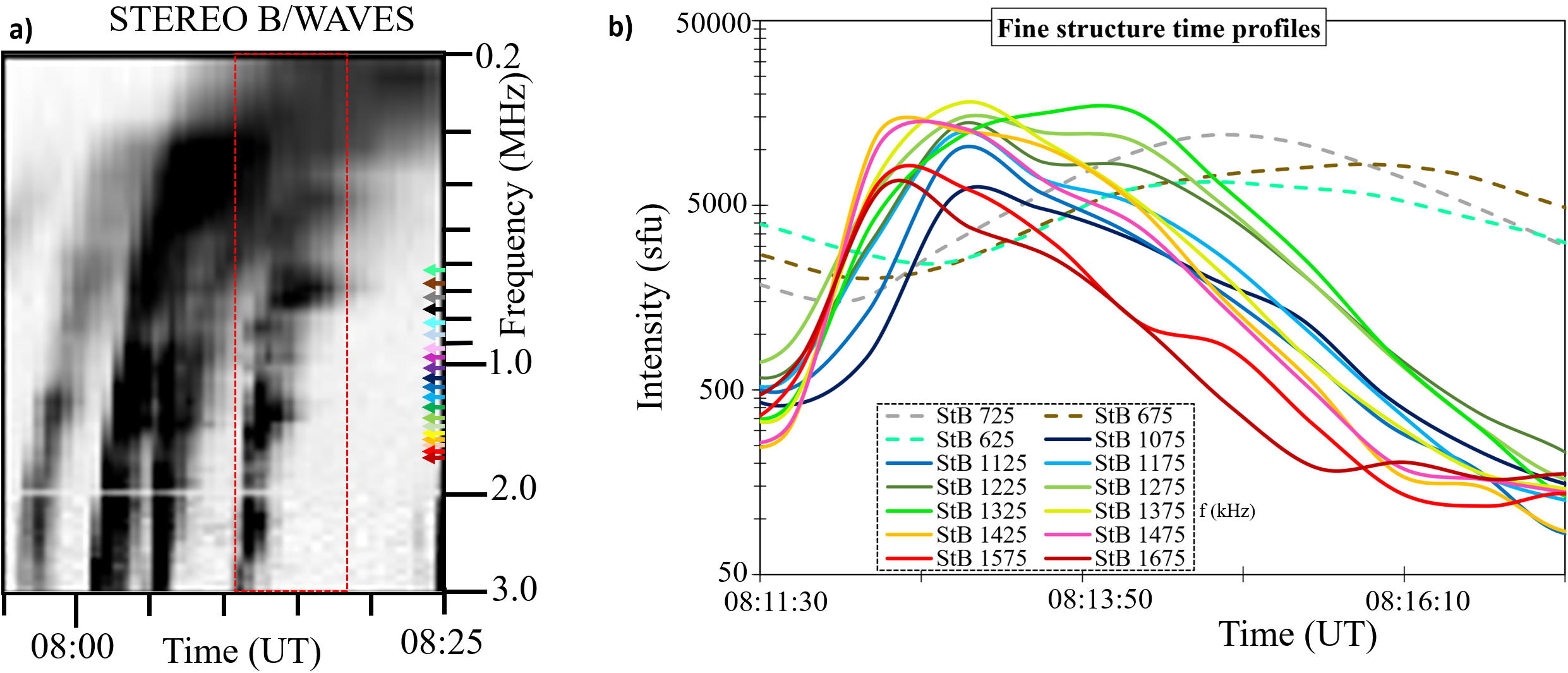}
  \end{subfigure}
  \caption{Spectral analysis of the structured type III observed on 12/11/2010. (a) The type III burst with triangular fine structures, observed at about 08:11~UT by the \textit{STEREO B} spacecraft. The considered frequency range is from 14 to 1~MHz. The dotted-red box represents the range where the time-profile analysis was performed. The colored arrows in the Y-axis represent the frequencies at which the single frequency time-profiles in the panel (b) were made . (b) The time profiles of triangular substructures.  The colors represent the same frequencies as marked in the panel (a)}
  \label{Fig:figure_7}
  \end{figure*}

We find the structured type III, observed at about 08:11~UT, very particular. The details on this radio burst are presented in Fig. \ref{Fig:figure_7} and \ref{Fig:figure_8}. Except being observed by the \textit{WAVES} instruments, this structured type III was also observed by the Nancay decametric array \citep[\textit{NDA};][]{Boischot80} at higher frequencies. Other bursts in the group (Fig. \ref{Fig:figure_6}) were either not in the time window of \textit{NDA} observations, or like in the case of last burst in the group, not observed. The \textit{NDA} observations of structured type III suggest that the source of the radio burst was located on the visible side of the Sun, as seen from Earth. The fast electron beam generating this type III probably propagated mostly in the direction of \textit{STEREO B/WAVES}, starting from the high latitudes with the trajectory allowing it to be observed by all three instruments.


\subsubsection{Time profiles for November 12, 2010 event} \label{2_time_profiles}

\begin{figure*}[h]
  \centering
  \begin{subfigure}[b]{0.575\textwidth}
    \includegraphics[width=\textwidth]{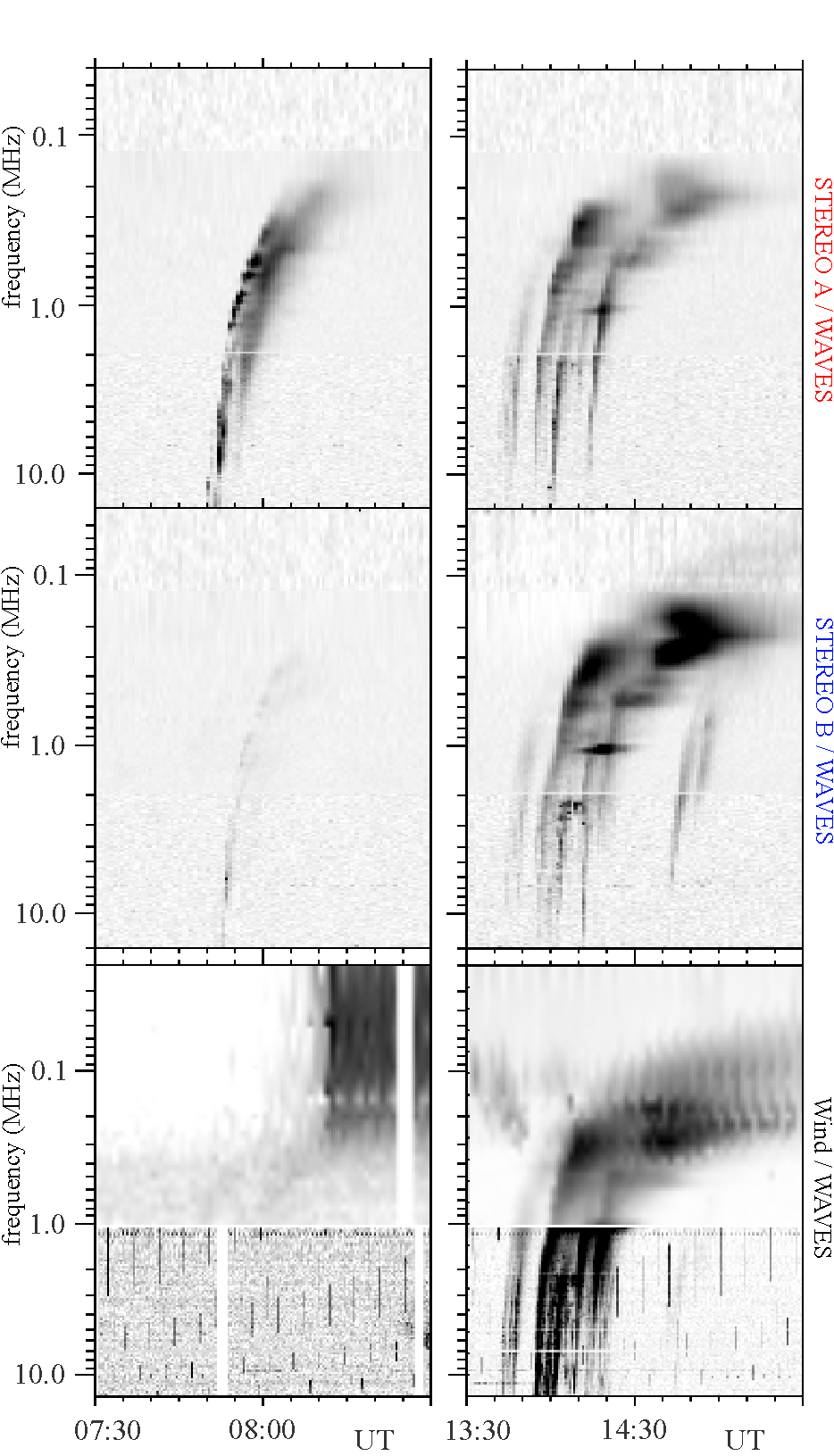}
  \end{subfigure}
  \caption{(a) Radio Dynamic spectra observed on 30/11/2011 between 7:30 UT and 8:30 UT. We note irregular type III with fine structures at 07:52 UT by the \textit{STEREO A} and \textit{STEREO B} spacecraft. (b) Radio dynamic spectra observed on 12/11/2010. We note striae like fine structures within the type III bursts observed at 13:45 UT, and large 'heart-shaped' substructures and striae bursts in the \textit{STEREO/WAVES} observations.}
  \label{Fig:figure_9}
  \end{figure*}

We performed the analysis of time profiles of the radio bursts in the similar way as in Sec. \ref{1_time_profiles}. The single frequency profiles of the triangular substructures of the type IIIb burst observed at 08:10~UT on 2/11/2010 are shown in Fig. \ref{Fig:figure_9}. We selected observations from \textit{STEREO B/WAVES} in which the intensity of this patchy burst (Fig \ref{Fig:figure_7}, panel (a) is highest. 

 \begin{figure}[h]
  \centering
  \begin{subfigure}[b]{0.49\textwidth}
    \includegraphics[width=\textwidth]{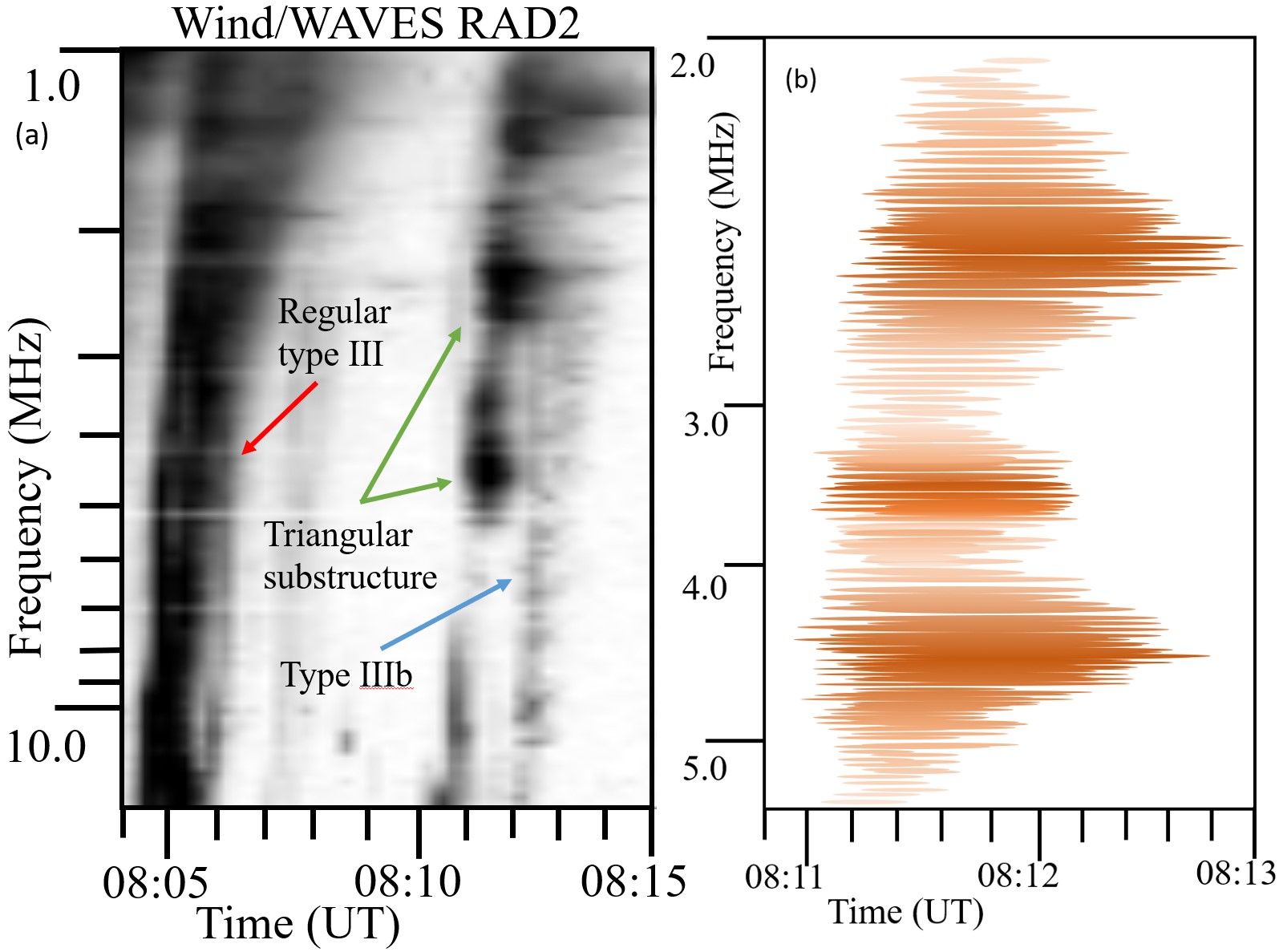}
  \end{subfigure}
  \caption{(a) The group of type III burst in the frequency range 14 to 1~MHz, observed by the \textit{Wind/WAVES}~RAD~2 instrument. The red arrow points to the regular type III observed at around 08:05~UT. The blue arrows point to the two structured type III bursts while the green arrow marks the periodic sub-structures observed at 08:11~UT. (b) A sketch of the structured type III burst observed at 08:11:00~UT and 08:13:00~UT.}
  \label{Fig:figure_8}
  \end{figure}
  

For the analysis of the time profiles, we select the triangular structure observed between 2 and 1~MHz, and the structure observed between 700 and 600~kHz. Fig. \ref{Fig:figure_7}a presents the zoomed in spectra, and the time profiles at frequencies which observe the bright triangular structure (Fig. \ref{Fig:figure_7}b). The envelope for this class of type III cannot be well distinguished due to overlapping fine structures. 


The time profiles of the triangular substructures overlap in rise part of a burst and their maximum is reached at the similar flux values. At 1075 and 1125~kHz, the full time profiles of the burst are shifted for about 5s forward.
The time profiles at frequencies above 1125~kHz have approximately simultaneous rise time and majority of them show the presence of a double peaked distribution of varying intensity, indicating possibly a fundamental and harmonic generation mechanism.

The time profiles of the triangular substructure observed at 725, 625 and 675~kHz (Fig. \ref{Fig:figure_7}b), and the structures observed at higher frequencies have comparable intensities but the morphology of the profiles is different. The profiles at three low frequencies are flatter than the one at frequencies above 1000~MHz, and their start time is shifted for about 10~s later. The flattening and the time shift of the profiles indicates the sudden decrease of the velocity of the electron beam responsible for generation of the radio emission \citep[see e.g.][]{Baselyan74a,Baselyan74b,Sawant78}
Generally, the drift rate of the structured type III is 13~kHz s$^{-1}$ i.e. $\approx 0.23$c. This drift rate is comparable to the one of the regular type III, observed during the same radio event (about 10~kHz s$^{-1}$ i.e. $\approx0.18$c.)

In order to investigate the complex maxima of the time profiles presented in Fig. \ref{Fig:figure_7} we also examine the \textit{Wind/WAVES}~RAD~2 observations in the frequency range 10--1~MHz. These observation have better spectral resolution (50 kHz and 200 kHz, respectively) and comparable temporal resolution to the \textit{STEREO/WAVES}~HFR2 spectra (20s and 16s, respectively). While \textit{STEREO} observations show one structured type III burst at about 08:11 UT (Fig. \ref{Fig:figure_7}), the \textit{Wind} observations show two structured bursts, at 08:10:30~UT and  08:12:00~UT (Fig. \ref{Fig:figure_8}a). It is possible that these two bursts are the fundamental and harmonic component, but from the available observations this is not clear. Both bursts are composed of a number of striae of varying intensity, and it seems that the dense and intense striae compose the  more broadband triangular structures of the type III bursts (Fig. \ref{Fig:figure_8}). The striae substructures are similar to those found composing the type IIIb bursts (Sec.~\ref{1_case_study}). The striae have bandwidth of approximately 30 -- 80 kHz, with the lowest measured bandwidth being in the limit of the resolution of the \textit{Wind/WAVES}~RAD~2  observations . These observations indicate that striae of lower bandwidth may also exist but they are not resolved. The duration of striae is about 30~s when presumably they do not form the triangular substructure, and 60~s when within the triangular structure. We note that the decay time of type III bursts in this frequency range is about 60~s.



\subsection{Type III bursts with irregular substructures}\label{3_irregular}

The last class of structured interplanetary type III bursts are the one with irregular fine structures.  Fig. \ref{Fig:figure_9} shows few examples of type III bursts with irregular substructures. As we can see, this class of type III bursts have fine structures which do not show a periodicity or regularity in their morphology. They are more seldom observed than other types of structured type III bursts, and their fine structures can strongly vary in bandwidth, duration and shape (Fig. \ref{Fig:figure_9}). The morphology of the envelope of type III radio bursts with irregular substructures is similar to the ordinary type III radio bursts. It was not possible to analyse such bursts in systematic way, and individual case studies are necessary to possibly draw some more general conclusions. 

Fig. \ref{Fig:figure_9}a shows an example of type III radio burst with irregular fine structures as observed by \textit{STEREO A/WAVES} on 30/11/2011. The intensity of the type III burst slowly decreases from about 10~MHz towards 2~MHz. Starting from about 1200~kHz, emission is well observed again, composed of dot-like fine structures within the type III envelope. The bandwidth of the dot-like fine structures is narrow ($\ll$ 40 -- 60 kHz), and duration seem to be significantly shorter that for the type III bursts at these frequencies. However, the structures are on the edge of the time/frequency resolution and it is not possible to reliably estimate their duration and bandwidth. The structured type III is subsequently followed by a faint and fuzzy type III burst without clearly distinguishable fine structures. These two bursts could be fundamental and harmonic components of the same burst. Similar fundamental-harmonic pair was recently reported in the decameter range \citep[][]{Zhang20}.

Another group of type III radio burst with irregular fine structures is presented in Fig. \ref{Fig:figure_9}b. This very particular radio event, observed on 12/11/2010 (from about 13:30 until 15:50~UT) was associated with a flare/CME event. Eruptive events are often strongly disturbing the ambient plasma conditions and can produce turbulence in the plasma environment through which the electron beams propagate. This can result in a large number of  fine structures observed in the associated radio emission \citep[e.g.][]{Magdalenic20}. The presented radio event also shows strongly fragmented radio emission. The event started at about 13:35~UT, with groups of highly structured type IIIb bursts observed by the \textit{Wind/WAVES}~RAD~2 receiver (13 -- 1~MHz). These bursts were very weak in both \textit{STEREO/WAVES} observations. At about 13:45~UT, a new group of structured type III bursts were observed with irregular substructures and this time, better observed in \textit{STEREO/WAVES} observations (Fig. \ref{Fig:figure_9}). The fine structures in type III bursts associated with this event were not isolated in \textit{Wind/WAVES} observations due to the overall higher intensity of the apparently overlapping radio bursts. The fine structures however, can be seen in \textit{STEREO A} \& \textit{B/WAVES} where the emission intensity is not as high as observed by \textit{Wind/WAVES}. Very particular fine structures are observed, in \textit{STEREO/WAVES} observations resembling large striae structures, and even heart-like shaped substructure (Fig. \ref{Fig:figure_9}). The bandwidth of striae-like fine structures is $\sim$ 100 kHz and their duration seem to increase with frequency decrease. The fine structures do not show any form of drift rate. These fine structures could be generated by the interaction between the electron beam and over-dense structures such as blobs in solar wind which are common in slow-solar wind \citep[][]{Sanchez17a}.  Another possible interpretation could be that the fine structures are enhancements at discrete frequencies of a regular type III burst source (electron beam) which propagates in a highly turbulent medium. 



\section{A model for striae emission}
\label{Sec:model}

It is difficult to explain all different variations of the type III fine structures presented in this study. Therefore, we focus on the basic ideas that may explain general features of structured bursts with the striae elements. We will do it employing detailed analysis of the radio event described in Sec.~\ref{1_case_study}. The generation and time evolution of the striae elements within type III radio bursts can be self-consistently described using a probabilistic model (PM) of the beam plasma interaction (BPI) in randomly inhomogeneous plasma \citep[][]{Krafft13, Krafft14, Krafft15, Krafft16, Krafft17, Voshchep15a, Voshchep15b, Krasnoselskikh19, Tkachenko21}. Herein, we present a scenario for the generation of striae elements described in Sec.~\ref{1_case_study}. We note that the same generation mechanism can be applied for the other type III fine structures presented in this study as well.

 \begin{figure*}[h]
  \centering
  \begin{subfigure}[b]{0.99\textwidth}
    \includegraphics[width=\textwidth]{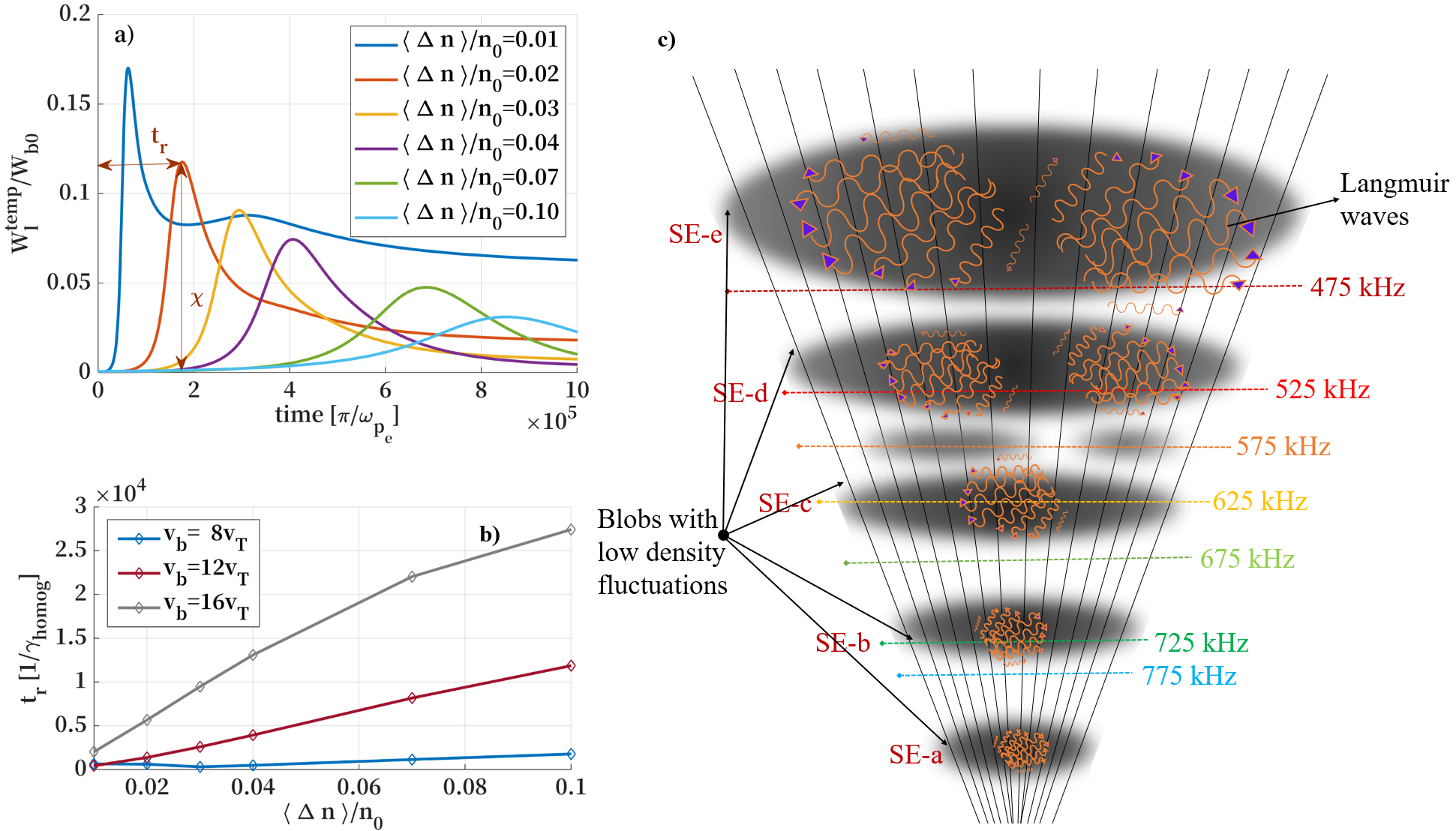}
  \end{subfigure}
  \caption{Effects of random density fluctuations on the growth of Langmuir waves that lead to electromagnetic emission. (a) Temporal evolution of the Langmuir wave intensity $W_l^{temp}$, in the units of electron beam energy $W_b$, is shown as a function of level of density fluctuations. The time of growth of the Langmuir wave till the maximum amplitude $x$ is denoted as $t_r$. (b) The time of growth of Langmuir wave energy ($t_r$, panel a) during the beam relaxation, in units of inverse growth rate for homogeneous plasma $t_r[1/\gamma_{homog}]$, is presented as a function of density fluctuations $\Delta n/n_o$. Three different electron beam speeds ($V_b$ in units of the thermal speed $V_T$; where $V_T$ is considered to be 0.01c) are considered. (c) The cartoon representation of an electron beam propagating through diverging magnetic field lines interacting with density blobs of different scales at different frequencies. The frequency levels are marked based on Fig. \ref{Fig:figure_5} and they roughly correspond to the different regions where the striae and envelope are observed. The enhanced growth of Langmuir waves is shown in different blobs with the wavy orange markers. We note that panels (a) and (b) were adapted from \cite{Tkachenko21}.}
  \label{Fig:PM_BPI}
  \end{figure*}

It is widely accepted that the generation of type III bursts is a two-step process \citep[see for e.g.][]{Ginzburg58, Suzuki85book, McLean85book}. Initially, the beam of fast electrons generates Langmuir waves which are then emitted in the form of electromagnetic waves. Therefore, the characteristics of the electromagnetic emission are determined by the characteristics of the electron beam and, by the processes of evolution of Langmuir wave spectra followed by radio emission. Since electrons are magnetized and move along the magnetic field lines, the increasingly larger emitting area corresponds to the diverging magnetic field lines. The emitting volume covers a large region that can be characterised by a rather wide range of plasma densities. Herein we propose that the density fluctuations determine the characteristics of the type III fine structures, i.e., striae elements, their variations in brightness, dynamics and in particular, the duration of the growth phase.

The density fluctuations are ubiquitous in the heliosphere and can be present in a wide range of scales \citep[][]{Neugebauer75, Celnikier83, Celnikier87}. We note that $\Delta n/n_o$ (average density fluctuations) changes based on length scales, and it may be in a very broad range. Considering 400 km gives a range 0.1 \-- 1 \% or less, but $\Delta n/n_o$ over larger distance e.g., 40,000 km could be 10 \% or more. We note that the correlation scales of density fluctuations can be determined only from direct in-situ measurement. These kinds of studies are rare \citep[see e.g.,][]{ChenC20}. The amplitude of the density fluctuations are not very large but are still important for the evolution of BPI. For this analysis we are particularly interested in fluctuations in the range of spatial scales between several hundred to several thousand times the wavelength of the Langmuir waves, up to the distances corresponding to the relaxation length of the beam in homogeneous plasma for each given frequency of wave. The relaxation length of the beam is defined as:

\begin{equation}
    L \sim \frac{ln \Lambda V_b}{\gamma}
\end{equation}
    
Here $V_b$ is the characteristic beam velocity, $\gamma$ is the characteristic increment of BPI, and $ln~\Lambda$ is the Coulomb logarithm determined by the macroscopic parameters of plasma ( $\Lambda \sim (n\lambda_{D}^{3})$, $n$ is the density and $\lambda_D$ is the Debye length). 



\cite{Krasnoselskikh19} have shown that the reflection of the Langmuir waves from the density inhomogeneities, even as low as 1\%, can lead to a partial transformation of the energy of Langmuir waves into electromagnetic, i.e. radio emission. Recent observations show that the density fluctuations can be of a small to moderate size,  and they are usually in the range of 2~--~3\%. However, density fluctuations as large as 7\% \citep[see for e.g.][]{Krupar20} at distances corresponding to hectometric frequencies, were also reported. The bandwidths of striae structures reported in our study correspond to the low levels of density fluctuations, in accordance to \cite{Krasnoselskikh19}. The upper level of density fluctuations reported by \citep{Krupar20} would result in a significantly large bandwidths of the striae elements.


The PM predicts two regimes of evolution of the Langmuir wave spectrum. In the first regime a level of the density fluctuations is low, the wave dispersion dominates and this results in the continues growth of the Langmuir waves, which is not observed in the case of type III radio bursts. The condition for this regime can be given by

\begin{equation}
    3k^2\lambda_D^2~>~(<\Delta n^2/n_o^2>)^{1/2}
\end{equation}

here $<...>$ denotes averaged values. The second regime is characterised by a two phase evolution of Langmuir waves, i.e., the initial growth is followed by the decay. The condition for the second regime is opposite to previous and is defined as:

\begin{equation}
    3k^2\lambda_D^2~<~(<\Delta n^2/n_o^2>)^{1/2}
\end{equation}

The generation of type III radio bursts occurs under conditions corresponding to the second regime when the Langmuir waves are generated by the beam with the phase velocity approximately equal to the beam velocity. In this case, when the beam propagates with a rather small angle to the background magnetic field, the dispersion term may be written as $k^2\lambda_D^2 \simeq {T_e}/{W_b}$ (where $T_e$ is the temperature of the background plasma, and $W_b$ is the characteristic energy of the beam).


In the case of beam energies of a few keVs to tens of keVs (overwhelming majority of the bursts), the interaction will take place in the second regime where it is valid $k^2\lambda_D^2 \simeq {T_e}/{W_b}$. The two different regimes of BPI are realized as a function of the ratio between two parameters: ${T_e}/{W_b}$ and $(<\Delta n^2/n_o^2>)^{1/2}$ in the range of the characteristic scales determined above. Fig.~\ref{Fig:PM_BPI}a shows the spectrum evolution when the density fluctuations are large enough to be dominant over the wave dispersion $({3T_e}/{W_b}~<~(<\Delta n^2/n_o^2>)^{1/2})$.

A high level of density fluctuations also determines the process of electromagnetic wave generation at the fundamental plasma frequency ($\omega_{pe}$). Namely, when the density fluctuations dominate over the wave dispersion effects, the dominant process is the direct transformation of a part of Langmuir wave energy into electromagnetic waves. This happens in the vicinity of the density humps where the reflection of the Langmuir waves occur \citep[][]{Krasnoselskikh19}. This process also affects the evolution of the electromagnetic wave intensity, at the place of its generation. The intensity of electromagnetic emission evolves similarly to the evolution of Langmuir wave intensity. In the region where the Langmuir waves are generated, the growth phase of the electromagnetic emission will be approximately the same as the growth phase of the Langmuir waves and will depend on the level of the density fluctuations as shown in Fig.~\ref{Fig:PM_BPI}.

The numerical solutions presented in Fig. \ref{Fig:PM_BPI}a were obtained under the conditions ${n_b}/{n_o} = 10^{-5}$, $V_b = 16V_T$ (here, we considered the thermal speed of electrons in the solar wind to be, $V_T$ = 0.01c; \cite{Helekas20}), and for six different levels of average density fluctuations $<{\Delta n}/{n_o}>$ (adapted from \cite{Tkachenko21}). The increase in the level of density fluctuations induces at the same time, an increase in the growth time of the wave and a decrease in the peak value of the wave amplitude.


Figure \ref{Fig:PM_BPI}b shows that for electron beams of different velocity, the time needed for the growth of Langmuir wave energy, during the beam's relaxation, is greatly affected by the level of background density fluctuations. The faster beams will need more time to relax when the levels of density fluctuations increases. These effects appear as a natural consequence when the electron beam propagates through regions of varying density. Figure~\ref{Fig:PM_BPI}c shows a sketch of regions with varying level of density fluctuations and intensity of radio emission. The shaded areas represent increasingly larger blob regions where the level of density fluctuations is low in comparison to ambient regions outside of it. In these areas the wave growth rate is high, the amplitude of the Langmuir waves is large and accordingly the intensity of radio emission is also larger. On the other hand, the nearby "envelope areas" correspond to higher level of the density fluctuations, smaller amplitude of Langmuir waves and consequently low intensity of radio emission. These regions correspond to the low intensity radio emission, i.e., the envelop of type IIIb.

 \begin{figure}[ht!]
  \centering
  \begin{subfigure}[b]{0.48\textwidth}
    \includegraphics[width=\textwidth]{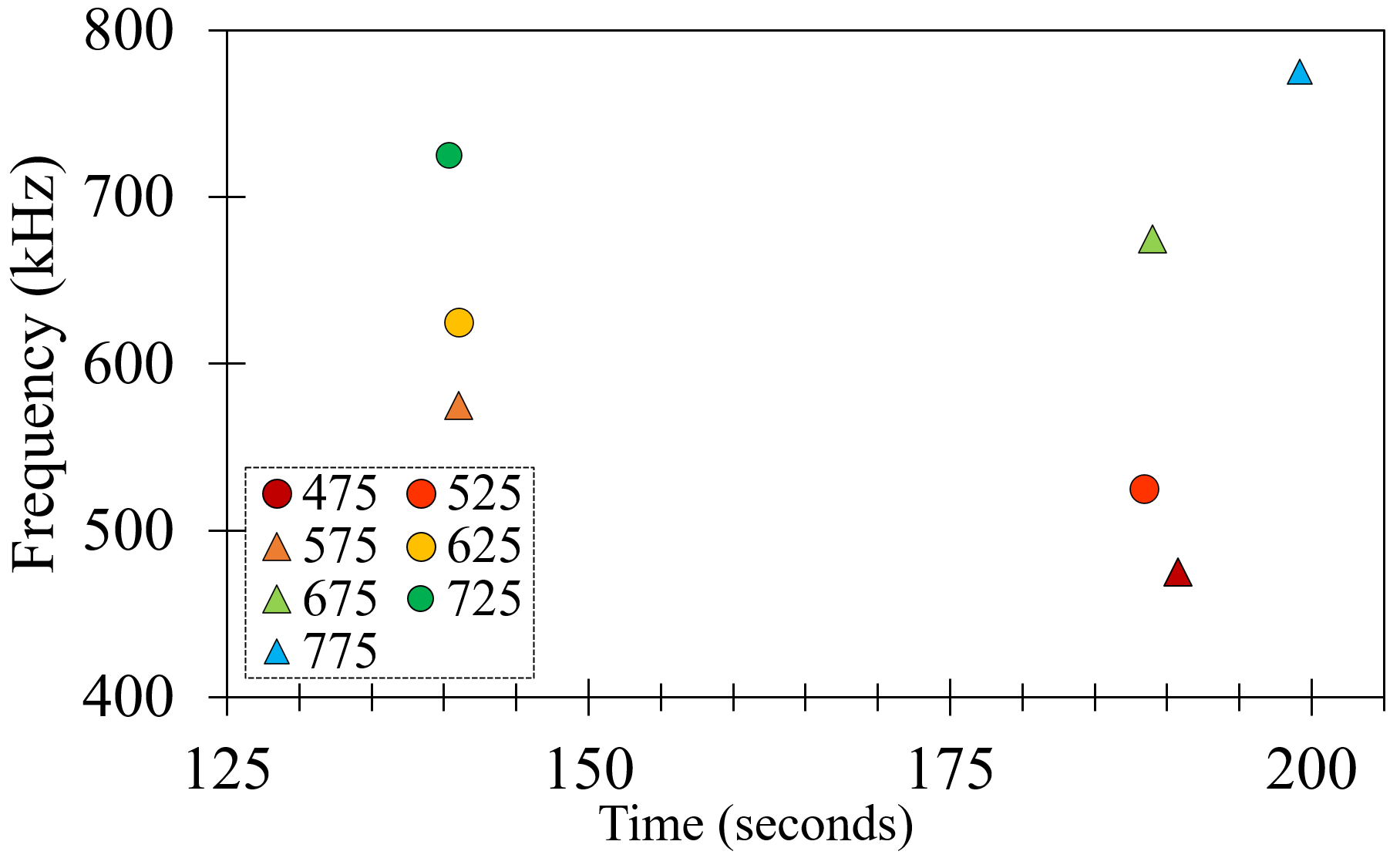}
  \end{subfigure}
  \caption{Rise times of the frequency-time profiles corresponding to the striae and their envelope in the frequency range of 800 -- 400 kHz. The triangle and circle markers represent the rise times of the envelope and striae elements, respectively.}
  \label{Fig:risetime}
  \end{figure}


We note that the times characterising BPI are not exactly the same as the characteristic times of evolution of the observed radio spectrum. Namely, depending on the characteristics of the ambient plasma, the radio emission can undergo different processes when propagating from the region where it was generated towards the observer.

When the plasma region where the radio emission is generated is optically thin, the electromagnetic waves propagate towards the observer with a small delay which is defined by the propagation time of light. When the source region of radio emission is optically thick, the initial stage of the growth phase of the wave intensity profile will be similar to the growth of the Langmuir waves. The reason for that is the rather fast transformation of the Langmuir waves into electromagnetic waves in the case of fundamental radio emission. The duration of the decay phase can be determined by different processes such as e.g. electromagnetic wave diffusion on the same density fluctuations \citep[e.g.][and references therein]{Arzner99, Bian19, Kontar19}. In this case, the decay phase of emission is determined by the characteristic time of electromagnetic wave diffusion in the regions close to the source region of the radio emission. This characteristic time may therefore increase with increasing levels of background density fluctuations.

Figure \ref{Fig:risetime} shows the rise times for the striae and envelope time profiles (see Fig. \ref{Fig:striae_envelop_tf_peak}a). The characteristic time of the rising phase of the burst $t_r$ (Fig. \ref{Fig:PM_BPI}b) can be estimated as,

\begin{equation}
    t_r \sim \alpha/\gamma_{lh}.
\end{equation}

In this relation, $\alpha$ is the coefficient characterising the dependence of the growth phase for different levels of density fluctuations (as presented in Fig. \ref{Fig:PM_BPI}a) and $\gamma_{lh}$ is a linear increment of BPI in homogeneous plasma that is defined as:

\begin{equation} \label{eq_5}
    \gamma_{lh} \simeq \omega_{pe} \frac{n_b}{n_o} (\frac{V_b}{\Delta V_b})^2
\end{equation}

In order to evaluate Eq.~\ref{eq_5} for the striae observed in Fig.~\ref{Fig:figure_5}, the characteristic rise times of the striae elements can be calculated assuming the following: $({V_b}/{\Delta V_b})\simeq10$,\ ${n_b}/{n_o}\simeq10^{-7} - 10^{-6}$,\ and $\omega_{pe}\simeq2\pi\cdot7\times10^{-5}$. We obtain $\gamma_{lh}\simeq0.5\times10^2$ seconds and this characteristic time corresponds to the typical parameters of the usually observed energetic electron fluxes, and is also in agreement with the estimates presented in Fig. \ref{Fig:risetime}.

\section{Discussion and Conclusions}\label{Discussions}

Fine structures of solar type III radio bursts can provide information about the ambient plasma conditions, such as e.g. coronal density and the magnetic field \citep[][]{Dulk78, Carley17}. Majority of studies of the radio fine structures, have so fare been performed at meter to decameter wavelengths \citep[see e.g.][]{Chernov07, Chernov11, Dabrovski20}. Recently, the new state-of-the-art instruments, LOFAR \citep[LOw Frequency ARray][]{Haarlem13} and MWA \citep[Murchison Widefield Array][]{Oberoi11}, provide us the high time/frequency resolution and high sensitivity observations that bring completely new insights in the particle acceleration and propagation processes \citep[see e.g.][]{Magdalenic20, Carley21}. 

Studies of fine structures observed in the hectometer and kilometer wavelengths are very rare, even in the \textit{STEREO} era, during which multi-point observations of the Sun are available. This work reports for the first time on the structured type III radio bursts in the space-based observations. We used observations of all three \textit{WAVES} instruments to confirm that the observed fine structures are not due the instrumental artifacts. It was not possible to use radio triangulation technique due to the not favourable spacecraft separation (around 180 degrees).


Based on their morphological characteristics, we can distinguish three main classes of structured interplanetary type III bursts: (a) the interplanetary type IIIb bursts, (b) type III with triangular substructures, and (c) type III bursts with irregular fine structures. The morphological similarities between metric to decametric and interplanetary type IIIb bursts indicate that the generation mechanism of the interplanetary type IIIb bursts could be inherently similar to that of the metric and decametric type IIIb bursts \citep[][]{Li12,Reid21Nat}. The elementary substructures of type IIIb bursts, in both frequency ranges, are named striae. We compared the type III fine structures observed by three different spacecraft, and using one-to-one schematic presentation  of the fine structures. This comparison allowed us to confirm that the type III fine structures are not due to instrumental effects.

Below we list some important results of the study.

\begin{itemize}
    \item [\textbullet]The time profiles of \textit{type IIIb bursts and their fine structures} showed that the striae structures have much higher intensity than the type IIIb envelope. The profiles of type IIIb burst are also more symmetrical than the profiles of the regular type III bursts. The striae structures have a symmetrical Gaussian-like profile which is significantly different from the double Gaussian like profiles of regular type III radio bursts.
\end{itemize}

\begin{itemize}
    \item [\textbullet]The \textit{Wind/WAVES}~RAD~2 observations show that \textit{triangular substructures of type III bursts} are possibly composed of a number of very narrowband striae structures (Fig.~\ref{Fig:figure_8}). This is probably the reason why the triangular substructures show complex morphology and time profiles.
    
\end{itemize}

\begin{itemize}
    \item[\textbullet]The general \textit{characteristics of type III substructures} are listed in the Table \ref{tab:my_table}. Additionally we also found that: a) the lower limit of the striae bandwidth seem to be imposed by the frequency resolution of observations; and b) the intensity enhancements observed within type III bursts are sometimes clearly composed of a number of striae structures (Fig.~\ref{Fig:figure_8}).
  \end{itemize}

Numerous studies reported type III fine structures in the metric to decametric wavelength range \citep[see e.g.][]{delaNoe76, Melnik19, Melnik21,Kontar17b, ChenX18, Sharykin18}. Some of these studies have also discussed the possibility that the structured type III bursts are precursor of the regular type III bursts and that the striae bursts, in the decametric type III-IIIb pair become more densely spaced towards the low frequency part of the bursts. This may be the reason why the striae bursts in the hectometric range appear to become quasi-continuous and are not well resolved in the presently available observations.

Recently \cite{Pulupa20} and \cite{ChenL21} presented observations of a hectometric type IIIb radio burst recorded by the Parker solar probe \citep[\textit{PSP};][]{Fox2016}. The clear individual striae bursts were observed close to 1 MHz, but not below that frequency. To our knowledge, all the studies up until now have only addressed observations of structured type III bursts in the metric to the low decametric wavelengths. This study, for the first time presents structured type III bursts in the hecto-kilometric wavelengths (interplanetary type IIIs), and opens the question about the cause of the observed structuring of radio emission in conditions of interplanetary plasma. We note that such well defined fine structures as the ones presented in this work are rarely observed. We have performed a survey of the interplanetary radio observations, taken by \textit{STEREO A}, \textit{STEREO B}, and \textit{Wind}, considering a two month period (01/09/2011 \-- 01/11/2011). We found a total number of about 710 type III radio bursts, out of which $\approx$ 45\% (310 bursts) were observed by all three spacecraft. Type III bursts with fine structures make about 16\% of the total type III bursts (118 bursts), out of which 41 bursts were observed by all three spacecraft.
Majority of structured type III bursts in the considered time interval were observed by one or two spacecraft. Furthermore, we found 15 cases (about 2\%) of type III bursts with very well defined fine structures (belonging to the first and second category defined in this study). On the other hand, irregular substructures (third category) were found far more frequently and accounted for the remaining 103 bursts. 


We have proposed an explanation for the generation of striae bursts employing the probabilistic model and simulations for the beam-plasma interactions \citep[][]{Krafft13, Krafft14, Krafft15, Krafft16, Krafft17, Voshchep15a, Voshchep15b, Krasnoselskikh19}. This widely discussed model can explain the generation of structured radio emission, regardless on the observing frequency range, assuming that the variations of the level of density fluctuations on different distances from the Sun result in variations of the average level of the wave amplitudes of the Langmuir waves. This in its turn results in different level of the electromagnetic emission. This phenomenon is taken into account in the PM of BPI and its effects on the different rise and decay phases of the evolution of BPI \citep[][]{Krafft13}. As a result, some of the source regions of the radio emission will be associated with large and some with small intensities of the radio emission. Observations confirming such a scenario are presented in Sec. \ref{Sec:model}.


If the striae bursts are generated due to the density-inhomogenities in the corona, they will also be dependent on the scale length of these density inhomogenities, as well as on other plasma parameters such as the ion and electron temperature \citep[][]{Ledenev04, Li12, Loi14}. Since the scales of density fluctuations in the interplanetary space are much larger than in the low corona \citep[see e.g.][]{Spangler02, Krupar20}, we can expect that the morphology of the striae in the interplanetary space will also reflect this difference. The herein presented observations indeed show that, the interplanetary type IIIb bursts have striae structures with somewhat larger bandwidths than the one in the metric range.


However, in order to be able to perform a more in-depth study of the type III fine structures, we need high resolution observations, preferably from two different point of view so the 3D location of the sources or radio emission can be estimated employing radio triangulation \citep[e.g.][]{Magdalenic14, Jebaraj20}. This additional information on the source positions of the radio emission will allow us to understand the type of ambient coronal structures through which the fast electron beams propagate. The observations from the novel missions, such as the Parker solar probe (\textit{PSP}), and Solar Orbiter \citep[\textit{SolO};][]{Muller13} can help us to better understanding the nature of the fragmented radio emission and the reason for its generation.

\begin{acknowledgements}
         The \textit{STEREO/SECCHI}  data  are  produced  by  a  consortium  of  RAL(UK), NRL(USA), LMSAL(USA), GSFC(USA), MPS(Germany), CSL(Belgium), IOTA(France), and IAS(France). The \textit{Wind/WAVES} instrument was designed and built as a joint effort of the Paris-Meudon  Observatory,  the  University  of  Minnesota,  and the  Goddard  Space  Flight  Center,  and  the  data  are  available at the instrument Website. We thank the radio monitoring service at LESIA (Observatoire de Paris) for providing value-added data that have been used for this study. I.C.J.\ was supported by a PhD grant awarded by the Royal Observatory of Belgium. I.C.J.\ and J.M.\ acknowledge funding by the  BRAIN-be (Belgian Research Action through Interdisciplinary Networks) project CCSOM (Constraining CMEs and Shocks by Observations and Modelling throughout the inner heliosphere), and BRAIN-be project SWiM (Solar Wind Modeling with EUHFORIA for the new heliospheric missions). V.K.(Vratislav Krupar)\ was supported by NASA grants 18-2HSWO2182-0010 and 19-HSR-192-0143. S.P.\ has received funding from the European Union’s Horizon 2020 research and innovation programme under grant agreement No 870405, and the projects C14/19/089  (C1 project Internal Funds KU Leuven), G.0D07.19N  (FWO-Vlaanderen), SIDC Data Exploitation (ESA Prodex-12).
\end{acknowledgements}



\bibliographystyle{aa}
\bibliography{bibtex27092012}
\end{document}